\documentclass[twocolumn]{aastex631} 

\begin{document}

\title{Brown Dwarf Candidates in the JADES and CEERS Extragalactic Surveys}

\author[0000-0003-4565-8239] {Kevin N.\ Hainline}
\affiliation{Steward Observatory, University of Arizona, 933 N. Cherry Ave, Tucson, AZ 85721, USA}

\author[0000-0003-4337-6211] {Jakob M.\ Helton}
\affiliation{Steward Observatory, University of Arizona, 933 N. Cherry Ave, Tucson, AZ 85721, USA}

\author[0000-0002-9280-7594] {Benjamin D.\ Johnson}
\affiliation{Center for Astrophysics $|$ Harvard \& Smithsonian, 60 Garden St., Cambridge MA 02138 USA}

\author[0000-0002-4622-6617] {Fengwu Sun}
\affiliation{Steward Observatory, University of Arizona, 933 N. Cherry Ave, Tucson, AZ 85721, USA}

\author[0000-0001-8426-1141] {Michael W.\ Topping}
\affiliation{Steward Observatory, University of Arizona, 933 N. Cherry Ave, Tucson, AZ 85721, USA}

\author[0000-0002-0834-6140] {Jarron M.\ Leisenring}
\affiliation{Steward Observatory, University of Arizona, 933 N. Cherry Ave, Tucson, AZ 85721, USA}

\author[0000-0003-0215-1104] {William M. Baker}
\affiliation{Kavli Institute for Cosmology, University of Cambridge, Madingley Road, Cambridge CB3 0HA, UK}
\affiliation{Cavendish Laboratory, University of Cambridge, 19 JJ Thomson Avenue, Cambridge CB3 0HE, UK}

\author[0000-0002-2929-3121] {Daniel J.\ Eisenstein}
\affiliation{Center for Astrophysics $|$ Harvard \& Smithsonian, 60 Garden St., Cambridge MA 02138 USA}

\author[0000-0002-8543-761X] {Ryan Hausen}
\affiliation{Department of Physics and Astronomy, The Johns Hopkins University, 3400 N. Charles St. Baltimore, MD 21218}

\author[0000-0002-4684-9005] {Raphael E.\ Hviding}
\affiliation{Steward Observatory, University of Arizona, 933 N. Cherry Ave, Tucson, AZ 85721, USA}

\author[0000-0002-6221-1829] {Jianwei Lyu}
\affiliation{Steward Observatory, University of Arizona, 933 N. Cherry Ave, Tucson, AZ 85721, USA}

\author[0000-0002-4271-0364] {Brant Robertson}
\affiliation{Department of Astronomy and Astrophysics, University of California, Santa Cruz, 1156 High Street, Santa Cruz CA 96054, USA}

\author[0000-0002-8224-4505] {Sandro Tacchella}
\affiliation{Kavli Institute for Cosmology, University of Cambridge, Madingley Road, Cambridge CB3 0HA, UK}
\affiliation{Cavendish Laboratory, University of Cambridge, 19 JJ Thomson Avenue, Cambridge CB3 0HE, UK}

\author[0000-0003-2919-7495] {Christina C.\ Williams}
\affiliation{NSF’s National Optical-Infrared Astronomy Research Laboratory, 950 North Cherry Avenue, Tucson, AZ 85719, USA}

\author[0000-0001-9262-9997] {Christopher N.\ A.\ Willmer}
\affiliation{Steward Observatory, University of Arizona, 933 N. Cherry Ave, Tucson, AZ 85721, USA}

\author[0000-0002-6730-5410] {Thomas L.\ Roellig}
\affiliation{NASA Ames Research Center, MS 245-6, Moffett Field, CA 94035, USA}

\begin{abstract}

By combining the JWST/NIRCam JADES and CEERS extragalactic datasets, we have uncovered a sample of twenty-one T and Y brown dwarf candidates at best-fit distances between 0.1 - 4.2 kpc. These sources were selected by targeting the blue 1$\mu$m - 2.5$\mu$m colors and red 3$\mu$m - 4.5$\mu$m colors that arise from molecular absorption in the atmospheres of T$_{\mathrm{eff}} < $ 1300K brown dwarfs. We fit these sources using multiple models of substellar atmospheres and present the resulting fluxes, sizes, effective temperatures and other derived properties for the sample. If confirmed, these fits place the majority of the sources in the Milky Way thick disk and halo. We observe proper motion for seven of the candidate brown dwarfs with directions in agreement with the plane of our Galaxy, providing evidence that they are not extragalactic in nature. We demonstrate how the colors of these sources differ from selected high-redshift galaxies, and explore the selection of these sources in planned large-area JWST NIRCam surveys. Deep imaging with JWST/NIRCam presents an an excellent opportunity for finding and understanding these ultracool dwarfs at kpc distances. 
\end{abstract}


\section{Introduction} \label{sec:intro}

At the intersection of stellar and exoplanet science lies the study of ultracool brown dwarfs, low-mass objects (M $< 0.07$ M$_{\odot}$) at relatively cool effective temperatures (T$_{\mathrm{eff}} <$ 2500 K). Over the last several decades researchers have uncovered many hundreds of brown dwarfs, which extend the stellar classification system from hot O-type stars down to L (T$_{\mathrm{eff}} = 1300 - 2000$K), T (T$_{\mathrm{eff}} = 700 - 1300$K), and Y (T$_{\mathrm{eff}} < 700$K) dwarfs \citep{kirkpatrick1999, cushing2011, kirkpatrick2021}. Selecting and characterizing these sources is vital for exploring the stellar/substellar initial mass function, understanding binary stellar evolution, and for increasing the census of stars around which potentially habitable planets may lie. 

However, at very low temperatures, brown dwarfs emit most of their flux in the infrared (IR) and are optically quite faint with spectral energy distributions (SEDs) that feature broad molecular and narrow resonance features (from H$_2$O, CH$_4$, NH$_3$, CO, and CO$_2$, among others). The observational depths for existing large area IR surveys restrict the recovered brown dwarfs to those in the Solar Neighborhood, although there are individual sources that have been found in the Milky Way thick disk and halo \citep{burgasser2003, loediu2010, murray2011, burningham2014, zhang2019, schneider2020}. It is vital to expand on these samples, as brown dwarfs at these distances help us understand the history of star formation in the Milky Way. 

Low-temperature brown dwarf candidates have been serendipitously discovered in deep near-IR surveys from the Hubble Space Telescope (HST), as they serve as a potential contaminant in the search for high-redshift galaxies and quasars \citep{bunker2004, wilkins2014, finkelstein2015, bouwens2015, banados2016}. Several authors have used the accumulated area of multiple deep HST near-IR surveys to perform systematic searches for brown dwarf candidates with observations out to $\sim 2 \mu$m \citep{ryan2011, masters2012, aganze2022}. Models of brown dwarf emission, however, show that while they do emit at 1 - 2$\mu$m, cold T and Y dwarfs peak at $4 - 5 \mu$m, outside of the wavelength coverage by HST, with faint fluxes requiring very deep observations. 

With the launch of the James Webb Space Telescope (JWST), a new chapter in finding low-mass stars and brown dwarfs has begun. The NIRCam instrument on JWST offers imaging across a FOV of 9.7 arcmin$^2$ at $0.7 - 5.0 \mu$m in the near-IR. Current and upcoming extragalactic surveys, which will cover tens to many hundreds of square arcminutes, will explore flux limits ideal for selecting cold brown dwarfs \citep{holwerda2018, hainline2020}. Indeed, multiple studies have predicted the number counts of brown dwarfs in JWST deep surveys \citep{ryanreid2016, aganze2022, casey2022}, and in the first year of JWST, there have been discoveries made of individual brown dwarf candidates in both the JWST Director’s Discretionary Early Release Science Program ERS 1324, GLASS, \citep{nonino2023} and the Cosmic Evolution Early Release Science (CEERS) Survey \citep{wang2023}. Excitingly, \citet{langeroodi2023} and \citet{burgasser2023} both independently confirmed a number of brown dwarf candidates using spectroscopy taken as part of the Ultra-deep NIRCam and NIRSpec Observations Before the Epoch of Reionization \citep[UNCOVER,][]{bezanson2022}, including the source from \citet{nonino2023}. These spectroscopic confirmations demonstrate the usage of deep extragalactic data for finding these rare sources. It should be noted that deep fields are chosen to point away from the disk of the Milky Way to avoid stellar contamination, and as a result the populations of brown dwarfs observed in these surveys may not be broadly applicable to the full Milky Way population. 

Following these studies, we aim to use the JWST Advanced Deep Extragalactic Survey (JADES) observations \citep{eisenstein2023} of the Great Observatories Origins Deep Survey (GOODS) Southern and Northern regions \citep{giavalisco2004} to explore the brown dwarf population at T$_{\mathrm{eff}} < 1300$K. These data, which we combine with deep ancillary HST observations at optical wavelengths, probe near-IR depths down to 29 - 30th magnitude AB. In addition, we supplement these observations with NIRCam observations from the CEERS Survey \citep{finkelstein2022} to increase the area covered at similar observational depths. Sources detected in these surveys offer a unique opportunity to understand brown dwarfs at kpc distances in the Milky Way thick disk and halo, and can be used to explore the types of subdwarf populations that will be observed in future very large area JWST surveys such as COSMOS-Web \citep{casey2022} and PRIMER \citep{dunlopprimer2021}. 

The outline of this paper is as follows. In Section \ref{sec:jades} we discuss the JADES and CEERS data we are using for finding brown dwarf candidates, the reduction of these data, and the extraction of photometry. We then describe the selection procedure for separating substellar candidates from galaxies. In Section \ref{sec:results} we present our full sample of 21 brown dwarf candidates, discuss our brown dwarf model fitting procedure, and the recovered properties of these objects, including their temperatures and distances. We also describe the measured proper motions for a subset of our sources. In Section \ref{sec:discussion} we explore the positions of these sources within the Galaxy, demonstrate methods to separate these sources from high-redshift galaxy candidates, and discuss the selection of brown dwarfs in other JWST deep extragalactic surveys. Finally, we conclude in Section \ref{sec:conclusion}. Throughout this paper we will be using magnitudes in the AB system \citep{oke1974}. 

\section{Data and Sample} \label{sec:jades}

\subsection{JWST NIRCam Observations} 

In order to explore the types of brown dwarf candidates that can be recovered in deep extragalactic fields, we looked at both the JADES and CEERS datasets, which consist of deep NIRCam observations in multiple filters spanning $1 - 5 \mu$m that we use to separate these sources from galaxies. These survey data have been explored both to find \citep{finkelstein2022, robertson2023,  hainline2023} and spectroscopically-confirm \citep{curtislake2023, haro2023, fujimoto2023} the farthest high-redshift galaxies as of the writing, making them ideal fields for understanding the role of brown dwarf contamination of extragalactic samples. 

\subsubsection{JADES}

JADES is an observing program that combines data from the NIRCam and NIRSpec extragalactic Guaranteed Time Observations (GTO) teams. The final data set includes NIRCam and MIRI imaging with NIRSpec spectroscopy across two well-studied fields: GOODS-S (RA = 53.126 deg, DEC = -27.802 deg) and GOODS-N (RA = 189.229, DEC = +62.238 deg). For this study, we focus on the NIRCam observations taken as of February 8 2023. The observations and reduction of the JADES data we use in this study are fully described in \citet{eisenstein2023}. Briefly, we use the JADES NIRCam observations taken as part of JWST PID 1180 and 1181 (PI:Eisenstein), as well as PID 1210 and 1286 (PI:Ferruit). 

In total, the JADES GOODS-S area is 67 square arcminutes, with 27 square arcminutes for the JADES Deep program, and 40 square arcminutes for the JADES Medium program. The filters used for the JADES Deep program in GOODS-S are NIRCam F090W, F115W, F150W, F200W, F277W, F335M, F356W, F410M, and F444W ($\lambda = 0.8 - 5.0 \mu m$), and the JADES Medium area was observed with the same filters, but without F335M. For the NIRCam parallel to the NIRSpec 1286 program, the F070W filter was included. The JADES GOODS-N area is 58 square arcminutes, split into a deeper southeast (SE) portion (27.6 square arcminutes) and a relatively shallower northwest (NW) portion (30.4 square arcminutes). For the GOODS-N observations, the filters used were F090W, F115W, F150W, F200W, F277W, F335M, F356W, F410M, and F444W, similar to JADES Deep in GOODS-S. 

We supplement the JADES observations with data from the publicly available JWST Extragalactic Medium Survey \citep[JEMS,][]{williams2023}, which focuses on the center of the JADES Deep region and uses NIRCam filters F182M, F210M, F430M, F460M, and F480M. In addition, we include F444W observations from the The First Reionization Epoch Spectroscopic COmplete Survey \citep[FRESCO, PID 1895,][]{oesch2023}. As part of this NIRCam grism survey of the GOODS-S and GOODS-N fields, NIRCam imaging was obtained in the F182M, F210M, and F444W filters. We combine the F444W imaging from the FRESCO data with the JADES imaging in the region where the surveys overlap. The 5$\sigma$ point-source depths in a $0.2^{\prime\prime}$ diameter circular aperture vary across the fields and filters between 28.8 and 30.6 AB Mag \citep[see][for more details]{eisenstein2023}. 

At shorter wavelengths, to uncover potential proper motions, we also use existing HST/ACS and WFC3 mosaics. We use the HST/ACS and HST/WFC3 mosiacs from the Hubble Legacy Fields (HLF) v2.0 for GOODS-S and v2.5 for GOODS-N \citep[$25' \times 25'$ for GOODS-S, and  $20.5' \times 20.5'$ for GOODS-N,][]{illingworth2013, whitaker2019}. We use data in the HST/ACS F435W, F606W, F775W, F814W, and F850LP filters to aid in exploring galaxy fits, and HST/WFC3 F105W, F125W, F140W, and F160W for exploring proper motions.

Details of the data reduction and flux extraction are described in \citet{robertson2023}, \citet{tacchella2023}, and \citet{eisenstein2023}. To select the brown dwarfs across the JADES survey, we use $0.2^{\prime\prime}$ diameter circular aperture magnitudes extracted from images that have been convolved to match the F444W point-spread function (PSF) to obtain accurate colors. We applied aperture corrections to these magnitudes using empirical JWST/NIRCam PSFs assuming point source morphologies for our sources \citep[see][for more details]{ji2023}. We note that the PSFs used in this analysis were derived from pre-flight pointing predictions. 

\subsubsection{CEERS}

To supplement our sample of brown dwarf candidates, we additionally explore The Cosmic Evolution Early Release Science (CEERS) Survey\footnote{https://ceers.github.io/} \citep{finkelstein2022} NIRCam observations. The CEERS Survey currently targets 90.8 square arcminutes of the Extended Groth Strip (EGS) with the NIRCam filters F115W, F150W, F200W, F277W, F356W, F410M, and F444W down to $5\sigma$ point-source depths of 28.4 - 29.3 AB Mag in a $0.2^{\prime\prime}$ diameter circular aperture. We combine the JWST data with public HST/ACS data covering the EGS in F606W and F814W (A. Koekemoer, private comm.). We used the reduction process outlined above on the raw CEERS observations, and again, chose the $0.2^{\prime\prime}$ diameter circular aperture magnitudes extracted from F444W PSF-convolved images to select sources. 

\subsection{Selecting Brown Dwarf Candidates}

The first of the brown dwarf candidates were found serendipitously as part of the search for high-redshift galaxies presented in \citet{hainline2023}. These sources were observed with red NIRCam F090W-F115W colors (similar to F090W Lyman dropout galaxies), blue F115W - F277W colors, and then exceedingly red F277W - F444W colors, producing a ``V''-shaped SED. In addition, these sources had point-like morphologies, often with diffraction spikes from the NIRCam PSF observed at long wavelengths. When we fits these sources with observed and simulated galaxy templates using the photometric redshift code EAZY \citep{brammer2008} following the procedure outlined in \citet{hainline2023}, the fits were generally very poor, with large $\chi^2$ results for the best fits. The galaxy templates that we used included both star-forming and quiescent sources, including those with very strong nebular continuum and line emission, and we allowed EAZY to combine all of the individual templates when fitting a given source. While galaxy templates could not reproduce the colors of these sources, they were quite similar to the very low temperature T dwarfs collected as part of the SpeX Prism Spectral Libraries\footnote{https://cass.ucsd.edu/~ajb/browndwarfs/spexprism/index.html}, as well as the simulated brown dwarf SEDs from \citet{marley2021} and \citet{karalidi2021}. 

To better understand these interesting brown dwarf candidates, we performed a search across the JADES and CEERS footprints based on the colors of the objects serendipitously discovered as well as the colors of brown dwarfs in the literature. In \citet{nonino2023}, the authors selected a brown dwarf candidate from the JWST GLASS survey with $m_{\mathrm{F444W, AB}} \leq 28$, F356W-F444W $\geq 1.5$ and F115W-F444W $\geq 2$, criteria that were chosen based on T and Y dwarf infrared colors presented in \citet{meisner2020}. For our search, we extended this selection down to $m_{\mathrm{F444W, AB}} \leq 28.5$ due to the increased depth of the JADES and CEERS survey data, and in addition, we explored two alternate color selection schemes: F277W-F444W $\geq 0.0$ combined with either F115W-F277W $\leq 1.0$ or F115W-F150W $\leq 1.0$. These two selection criteria were designed to probe the ``V''-shaped SED of these brown dwarf candidates, and were chosen based on the positions of the JADES galaxies and the colors of the \citet{marley2021} and \citet{karalidi2021} substellar atmospheric models. 

We find 134 unique sources are selected by any one of the three color selection criteria in GOODS-S (2.0/arcmin$^2$), 173 (3.0/arcmin$^2$) in GOODS-N, and 355 (3.9/arcmin$^2$) across the CEERS survey data. The increased density of selected sources in GOODS-N and CEERS is due to the decreased observational depth, as for shallower surveys, noise preferentially scatters sources into the selection boxes. Additionally, we observe an increased number of hot pixels in the GOODS-N and CEERS datasets, often seen in the NIRCam LW channel, boosting the observed F444W flux. 

\begin{splitdeluxetable*}{lccr|rrrrBl|rrrrr}
\tabletypesize{\scriptsize}
\tablecolumns{12}
\tablewidth{0pt}
\tablecaption{Brown Dwarf Candidate {\tt forcePho} Fluxes (nJy) \label{tab:bd_fluxes}}
\tablehead{
 \colhead{Object ID} & \colhead{RA} & \colhead{DEC} & \colhead{$r_{\mathrm{half}}$($\prime\prime$)} & \colhead{F090W} & \colhead{F115W} & \colhead{F150W} & \colhead{F200W} & \colhead{Object ID} & \colhead{F277W}
 & \colhead{F335M} & \colhead{F356W} & \colhead{F410M} & \colhead{F444W}}
 \startdata
JADES-GS-BD-1 & \phantom{1}53.026024 & -27.867171 & 0.001 & $37.06 \pm 1.85$ & $267.43 \pm 13.37$ & $139.33 \pm 6.97$ & $92.14 \pm 4.61$ & JADES-GS-BD-1 & $51.18 \pm 2.56$ & -  & $131.92 \pm 6.6$ & $427.53 \pm 21.38$ & $327.93 \pm 16.4$ \\
JADES-GS-BD-2 & \phantom{1}53.036286 & -27.883739 & 0.001 & $18.83 \pm 1.09$ & $115.93 \pm 5.8$ & $67.78 \pm 3.39$ & $41.41 \pm 2.07$ & JADES-GS-BD-2 & $23.72 \pm 1.19$ & $19.04 \pm 0.95$ & $65.16 \pm 3.26$ & $178.97 \pm 8.95$ & $129.29 \pm 6.46$ \\
JADES-GS-BD-3 & \phantom{1}53.056307 & -27.868408 & 0.002 & $0.35 \pm 0.6$ & $11.36 \pm 0.58$ & $4.81 \pm 0.45$ & $2.27 \pm 0.51$ & JADES-GS-BD-3 & $-0.01 \pm 0.01$ & -  & $6.52 \pm 0.59$ & $26.35 \pm 1.74$ & $19.67 \pm 1.06$ \\
JADES-GS-BD-4 & \phantom{1}53.076738 & -27.889947 & 0.003 & $-1.46 \pm 0.01$ & $1.91 \pm 0.47$ & $1.76 \pm 0.51$ & $-0.16 \pm 0.01$ & JADES-GS-BD-4 & $-0.23 \pm 0.01$ & -  & $3.59 \pm 0.6$ & $21.73 \pm 1.24$ & $24.04 \pm 1.2$ \\
JADES-GS-BD-5 & \phantom{1}53.084040 & -27.839348 & 0.001 & $0.09 \pm 0.16$ & $2.99 \pm 0.15$ & $1.91 \pm 0.13$ & $1.25 \pm 0.17$ & JADES-GS-BD-5 & $8.77 \pm 0.44$ & $2.17 \pm 0.47$ & $60.25 \pm 3.01$ & $482.67 \pm 24.13$ & $624.61 \pm 31.23$ \\
JADES-GS-BD-6 & \phantom{1}53.103912 & -27.908507 & 0.002 & $3.24 \pm 0.66$ & $25.58 \pm 1.28$ & $13.02 \pm 0.67$ & $5.66 \pm 0.59$ & JADES-GS-BD-6 & $4.25 \pm 0.81$ & -  & $15.69 \pm 0.84$ & $55.39 \pm 2.77$ & $43.55 \pm 2.18$ \\
JADES-GS-BD-7 & \phantom{1}53.105547 & -27.815154 & 0.001 & $12.98 \pm 0.94$ & $118.67 \pm 5.93$ & $58.75 \pm 2.94$ & $33.5 \pm 1.68$ & JADES-GS-BD-7 & $20.3 \pm 1.28$ & -  & $66.83 \pm 3.34$ & $242.62 \pm 12.13$ & $209.54 \pm 10.48$ \\
JADES-GS-BD-8 & \phantom{1}53.133756 & -27.825521 & 0.003 & $-1.09 \pm 0.01$ & $4.14 \pm 0.39$ & $0.86 \pm 0.44$ & $0.14 \pm 0.39$ & JADES-GS-BD-8 & $0.96 \pm 0.54$ & $-1.23 \pm 0.01$ & $6.52 \pm 0.54$ & $47.2 \pm 2.36$ & $49.2 \pm 2.56$ \\
JADES-GS-BD-9 & \phantom{1}53.161140 & -27.809163 & 0.001 & $1.4 \pm 0.32$ & $17.11 \pm 0.86$ & $9.1 \pm 0.46$ & $3.63 \pm 0.34$ & JADES-GS-BD-9 & $3.0 \pm 0.32$ & $1.72 \pm 0.62$ & $13.81 \pm 0.69$ & $53.26 \pm 2.66$ & $44.22 \pm 2.35$ \\
JADES-GS-BD-10 & \phantom{1}53.161800 & -27.831613 & 0.001 & $305.8 \pm 15.29$ & $2330.96 \pm 116.55$ & $1313.29 \pm 65.66$ & $934.8 \pm 46.74$ & JADES-GS-BD-10 & $559.6 \pm 27.98$ & $344.54 \pm 17.23$ & $1317.24 \pm 65.86$ & $4082.01 \pm 204.1$ & $2994.17 \pm 149.71$ \\
JADES-GS-BD-11 & \phantom{1}53.200319 & -27.764118 & 0.002 & $3.24 \pm 0.87$ & $34.55 \pm 1.73$ & $20.0 \pm 1.0$ & $5.95 \pm 0.71$ & JADES-GS-BD-11 & $7.55 \pm 0.47$ & $4.07 \pm 1.13$ & $50.98 \pm 2.55$ & $206.26 \pm 10.31$ & $172.51 \pm 8.63$ \\
JADES-GN-BD-1 & 189.036189 & \phantom{-}62.234159 & 0.003 & $5.21 \pm 1.74$ & $40.12 \pm 2.17$ & $16.42 \pm 1.7$ & $8.83 \pm 1.18$ & JADES-GN-BD-1 & $6.95 \pm 1.34$ & -  & $33.03 \pm 1.65$ & $185.23 \pm 9.26$ & $183.02 \pm 9.15$ \\
JADES-GN-BD-2 & 189.118004 & \phantom{-}62.268968 & 0.002 & $3.89 \pm 1.25$ & $26.68 \pm 1.33$ & $12.26 \pm 1.08$ & $4.53 \pm 0.99$ & JADES-GN-BD-2 & $3.04 \pm 1.13$ & $-2.09 \pm 0.01$ & $18.03 \pm 1.16$ & $86.01 \pm 4.3$ & $69.02 \pm 3.45$ \\
JADES-GN-BD-3 & 189.204787 & \phantom{-}62.245553 & 0.003 & $4.01 \pm 1.48$ & $74.93 \pm 3.75$ & $33.08 \pm 1.65$ & $14.5 \pm 1.23$ & JADES-GN-BD-3 & $17.99 \pm 1.22$ & $8.86 \pm 1.96$ & $107.81 \pm 5.39$ & $643.26 \pm 32.16$ & $638.52 \pm 31.93$ \\
CEERS-EGS-BD-1 & 214.767936 & \phantom{-}52.845299 & 0.004 & -  & $8.97 \pm 0.83$ & $3.53 \pm 0.97$ & $0.59 \pm 0.9$ & CEERS-EGS-BD-1 & $0.4 \pm 1.24$ & -  & $7.22 \pm 1.19$ & $31.04 \pm 2.97$ & $25.53 \pm 2.64$ \\
CEERS-EGS-BD-2 & 214.823254 & \phantom{-}52.874531 & 0.004 & -  & $8.89 \pm 0.88$ & $4.76 \pm 0.98$ & $1.81 \pm 0.83$ & CEERS-EGS-BD-2 & $1.12 \pm 1.0$ & -  & $10.52 \pm 1.28$ & $61.99 \pm 3.18$ & $50.82 \pm 2.54$ \\
CEERS-EGS-BD-3 & 214.851340 & \phantom{-}52.799294 & $<0.001$ & -  & $289.62 \pm 14.48$ & $138.55 \pm 6.93$ & $78.11 \pm 3.91$ & CEERS-EGS-BD-3 & $61.89 \pm 3.09$ & -  & $143.33 \pm 7.17$ & $478.62 \pm 23.93$ & $389.8 \pm 19.49$ \\
CEERS-EGS-BD-4 & 214.910274 & \phantom{-}52.860077 & 0.002 & -  & $81.6 \pm 4.08$ & $38.49 \pm 1.92$ & $29.99 \pm 1.5$ & CEERS-EGS-BD-4 & $14.22 \pm 0.74$ & -  & $35.32 \pm 1.8$ & $146.27 \pm 7.31$ & $115.95 \pm 5.8$ \\
CEERS-EGS-BD-5 & 214.913696 & \phantom{-}52.895919 & 0.002 & -  & $17.05 \pm 0.85$ & $8.55 \pm 1.07$ & $6.54 \pm 0.94$ & CEERS-EGS-BD-5 & $2.7 \pm 1.08$ & -  & $8.89 \pm 1.02$ & $34.09 \pm 2.81$ & $22.37 \pm 2.45$ \\
CEERS-EGS-BD-6 & 214.940850 & \phantom{-}52.907711 & 0.003 & -  & $3.28 \pm 0.64$ & $2.69 \pm 0.92$ & $0.05 \pm 0.68$ & CEERS-EGS-BD-6 & $0.21 \pm 1.29$ & -  & $4.63 \pm 0.93$ & $28.56 \pm 3.13$ & $33.92 \pm 2.56$ \\
CEERS-EGS-BD-7 & 214.951604 & \phantom{-}52.913456 & 0.003 & -  & $13.67 \pm 0.68$ & $7.38 \pm 1.0$ & $3.79 \pm 0.69$ & CEERS-EGS-BD-7 & $2.9 \pm 1.51$ & -  & $8.6 \pm 1.0$ & $60.33 \pm 3.7$ & $59.95 \pm 3.0$ \\
\enddata
\end{splitdeluxetable*}

After making our photometry and color cuts, we examined the resulting samples of objects and rejected non-astrophysical sources, including hot pixels in the F444W mosaic and diffraction spikes from bright stars in the footprint. We also rejected sources with obvious extended morphologies, as those are likely galaxies. To further aid in removing potential galaxies, our candidate selection also required that the {\tt EAZY} galaxy fit had $\chi^2 > 20$. One source, CEERS-EGS-BD-5, has an {\tt EAZY} galaxy fit with $\chi^2 = 18.31$, but we included this source in our final sample as the {\tt EAZY} fit is not consistent with the non-detections in the ACS F606W and F814W filters. We rejected those sources with red long-wavelength slopes similar to those described in \citet{furtak2022, endsley2022, barro2023, labbe2023a, labbe2023b, matthee2023, hainline2023}, as these have colors inconsistent with brown dwarf models. 

Our final sample consists of 11 sources in GOODS-S, 3 sources in GOODS-N, and 7 sources in CEERS, for a total of 21 brown dwarf candidates. For these candidates, we re-fit the photometry using {\tt forcepho} (BDJ in prep), a modeling code that combines S\'ersic profiles to measure the pixel-level fluxes for a given source directly from the individual NIRCam exposures. The usage of {\tt forcepho} on NIRCam data is described in \citet{robertson2023}.The {\tt forcepho} photometry agrees with the photometry used to select the original sources to within $3 - 26$\% varying between the filters (the F410M fluxes are the most discrepant, with the {\tt forcepho} fluxes being systematically brighter than those measured with a circular aperture), and none of the {\tt forcepho} colors for these sources are outside the selection criteria we used to initially select them. We impose a SNR $< 20$ limit on the {\tt forcepho} uncertainties to account for systematics. 

In \citet{wang2023}, the authors present a brown dwarf candidate selected from CEERS public data. In our independent reduction of the same dataset, we extracted flux for this source and found that the photometric measurements presented in their Table \ref{tab:bd_fluxes} are highly discrepant from our measurements. They present an AB magnitude in the F277W filter for this source of 29.02, while we measure a $0.2^{\prime\prime}$ diameter circular aperture magnitude in the same filter of $27.53 \pm 0.04$. A separate photometric catalog from the The DAWN JWST Archive\footnote{https://dawn-cph.github.io/dja/} for the source includes a $0.36^{\prime\prime}$ diameter circular aperture magnitude of $27.37 \pm 0.05$. In addition, the NIRCam colors of the source as presented in \citet{wang2023} are different from what we measure - we would not select this source based on our color criteria previously described. We are confident, based on the provided RA and DEC and the individual filter thumbnails presented in \citet{wang2023} for the source, that we were comparing to the same object within our own reduction of the catalog. The {\tt EAZY} fit to our measured fluxes for this object results in a fit at a photometric redshift $z_a = 5.9$ with $\chi^2 = 2.4$. Given that in our own reduction of these data this source would not be selected as a brown dwarf candidate, we do not include it in our analysis. 

\begin{figure*}
  \centering
  \includegraphics[width=0.98\textwidth]{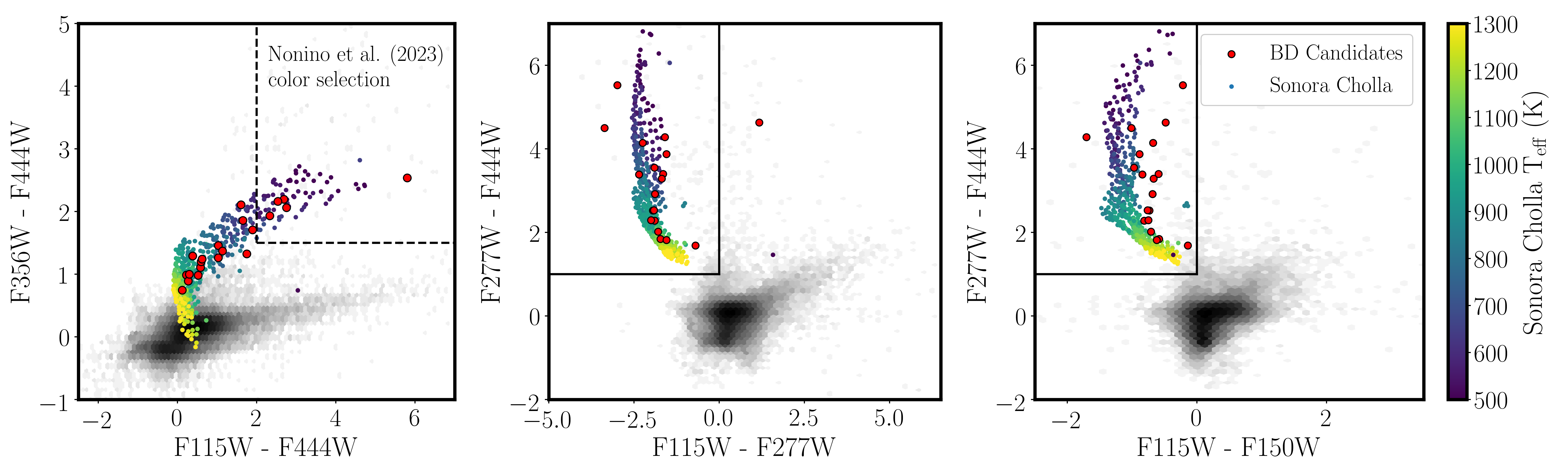}
  \caption{Color criteria used to select brown dwarf candidates in this work. In each panel, JADES sources with m$_{F444W} < 28.5$ are plotted in grey. Our selected JADES and CEERS candidates discussed in this paper are shown with red circles with black outlines, and the predicted ultracool dwarf colors from the Sonora Cholla models are plotted with small points colored by the T$_{\mathrm{eff}}$ values as shown in the colorbar. The black lines represent the color limits used for selecting sources. \textit{Left:} F356W - F444W color plotted against F115W - F444W color, with dashed black lines showing the color criteria adopted by \citep{nonino2023} for selecting their brown dwarf candidate in the JWST GLASS data. These colors would limit the resulting Sonora Cholla brown dwarf models to those with T$_{\mathrm{eff}} < 700$K. \textit{Middle:} F277W - F444W color plotted against F115W - F277W color, with the color limits used in selecting sources represented with solid lines. \textit{Right:} F277W - F444W color plotted against F115W - F150W color, with the color limits used in selecting sources represented with solid lines. Both the middle and right panels feature new color criteria designed to recover the ``V''-shaped SED seen in brown dwarf models from \citet{marley2021} and \citet{karalidi2021}.
  \label{fig:BD_Color_Color_All}}
\end{figure*}

In Figure \ref{fig:BD_Color_Color_All}, we plot the color selection regions that we used for selecting our candidates. In each panel, we represent the colors for the JADES sources measured using the $0.2^{\prime\prime}$ diameter circular aperture magnitudes with a greyscale density plot. The final sample of JADES and CEERS candidates is plotted with red points. We compare our candidates to the colors for the ``Sonora Cholla'' brown dwarf model atmospheres from \citet{karalidi2021}. We plot the solar metallicity Sonora Cholla points with colors shaded by the model effective temperature. We find that most of our sources would not be selected by the color criteria used in \citet{nonino2023}, as shown in the left panel of Figure \ref{fig:BD_Color_Color_All}. These color limits would only select Sonora Cholla models with effective temperature T$_{\mathrm{eff}} < 700$K (Y dwarfs). The selection criteria depicted in the middle and right panels of Figure \ref{fig:BD_Color_Color_All}, on the other hand, includes the majority of the brown dwarf Sonora Cholla model points across the full range of modeled T$_{\mathrm{eff}}$ values, and separates these sources from the bulk of the JADES galaxies. Our candidates have colors that nicely agree with the Sonora Cholla model colors, although one object, JADES-GS-BD-5, lies outside of the F277W-F444W vs. F115W-F277W selection box by virtue of its red F115W - F277W color. This source will be discussed further in the next section.  

To estimate the brown dwarf properties, including the distances to these candidates, we fit the NIRCam {\tt forcepho} fluxes for the sources with the \citet{karalidi2021} Sonora Cholla models. We do not fit the HST ACS or WFC3 fluxes because of the possibility of proper motion making it difficult to assure that the measured flux corresponds to the correct source (see the next section for a discussion of the sources with observed proper motion). The models are provided with three free parameters: effective temperature (T$_{\mathrm{eff}}$), surface gravity (log $g$), and eddy diffusion parameter (log $K_{zz}$). We compared each provided model against the observed {\tt forcepho} photometry, and varied the overall normalization of the model for a fourth free parameter. The adopted best fit was the one that resulted in the minimum $\chi^2$ for each of our candidates. 

We additionally fit the candidates using the ATMO 2020 solar metallicity brown dwarf atmosphere and evolutionary models from \citet{phillips2020}. These were generated using a 1D-2D atmospheric modeling code designed to study hot Jupiters and brown dwarfs, and the ATMO 2020 release includes both atmospheric and evolution models. For simplicity, we adopted the atmospheric chemical equilibrium models with only two free parameters: effective temperature and surface gravity. We performed a similar fit using these models as was done with the Sonora Cholla models, where the best fit was the one corresponding to the minimum $\chi^2$. 

\section{Results} \label{sec:results}

\subsection{Physical Properties} 

\begin{figure*}
  \centering
  \includegraphics[width=0.49\textwidth]{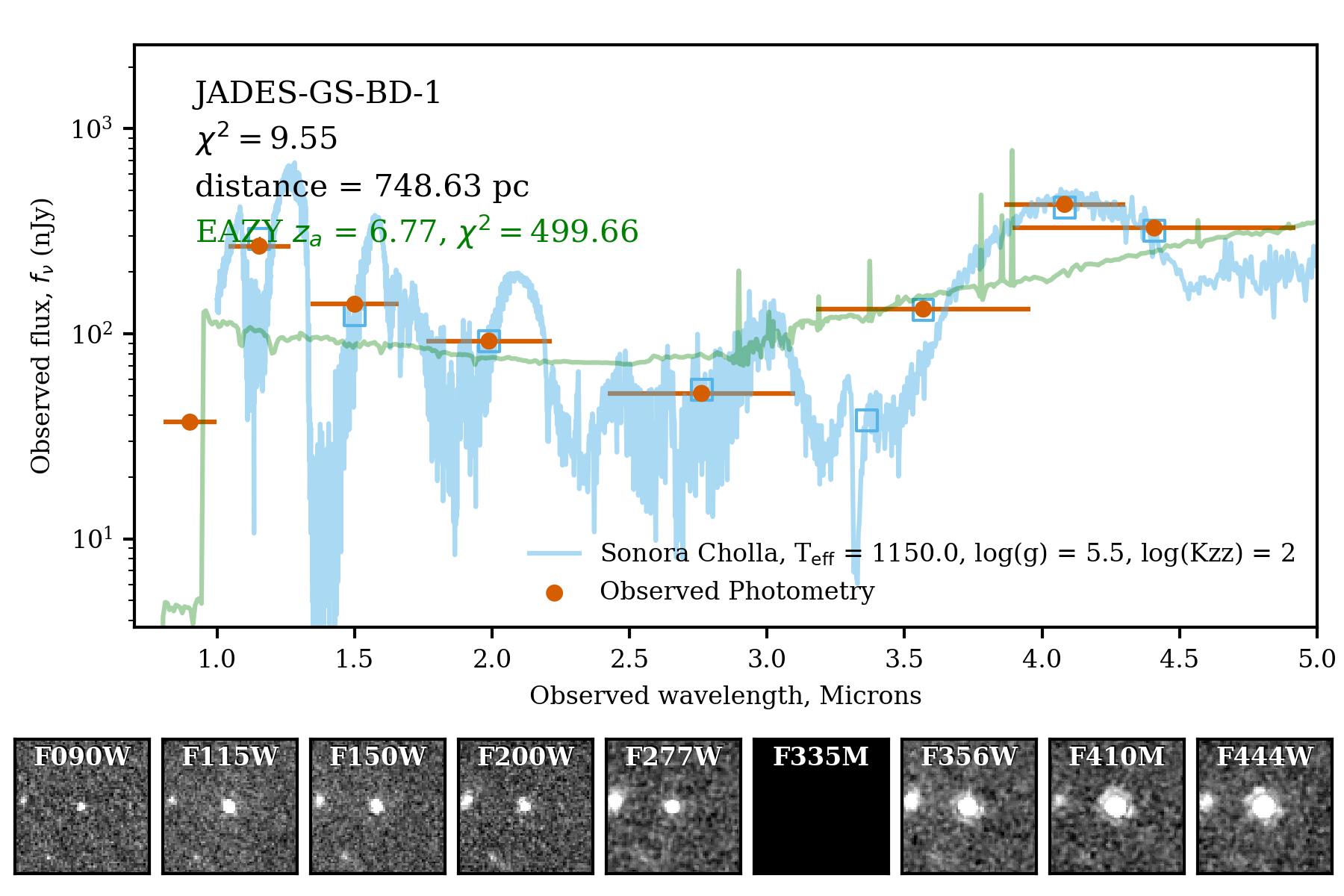}
  \includegraphics[width=0.49\textwidth]{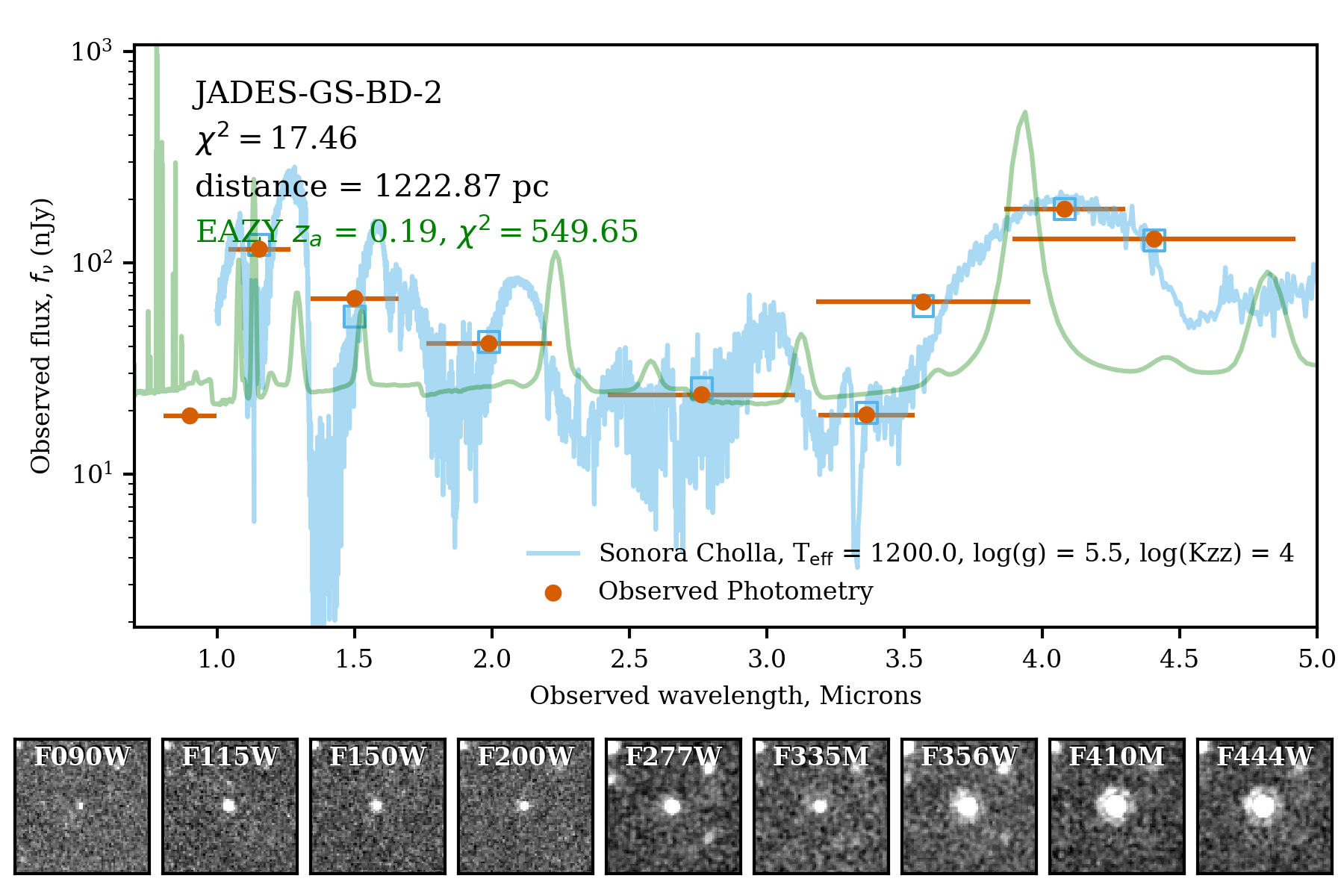}
  \includegraphics[width=0.49\textwidth]{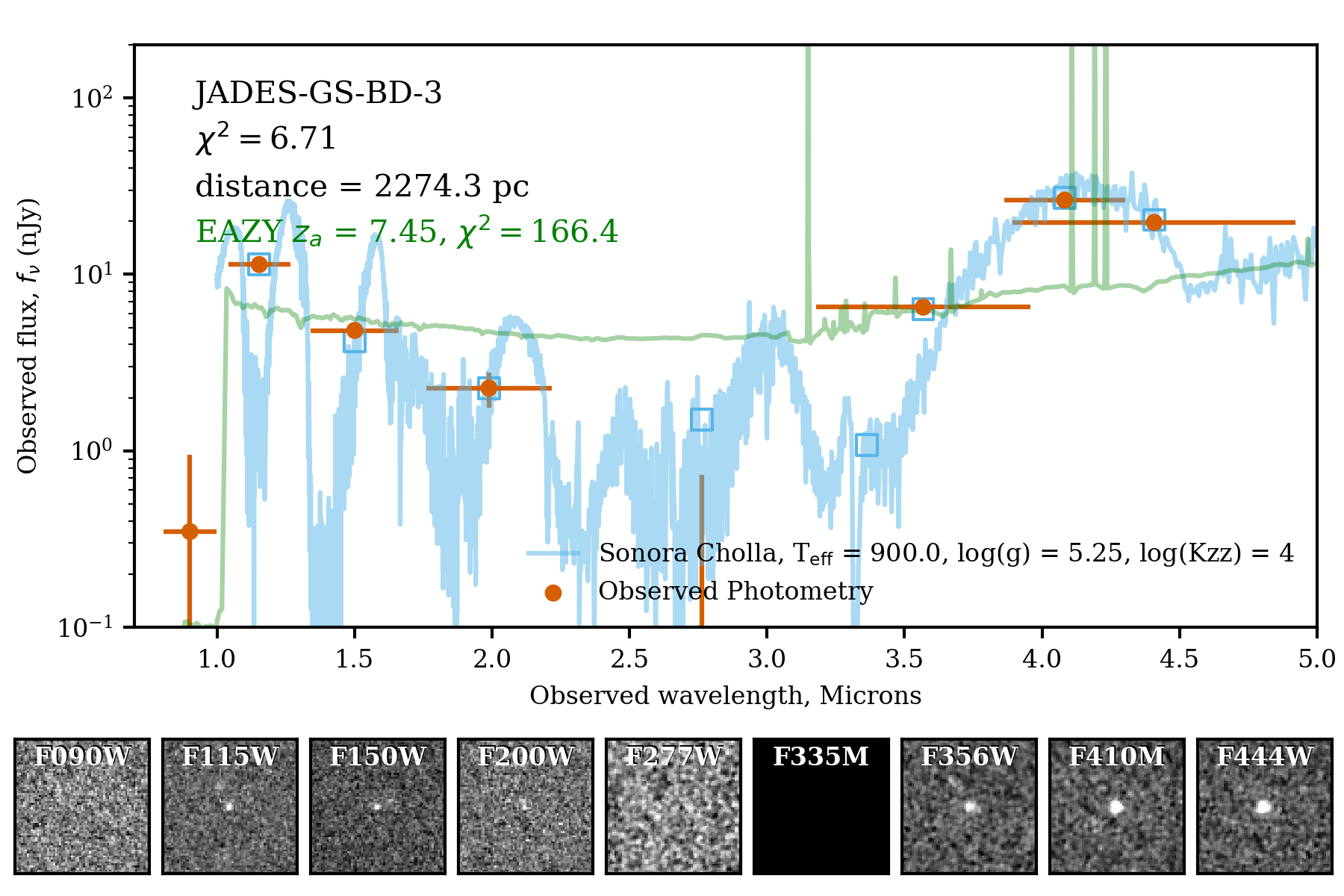}
  \includegraphics[width=0.49\textwidth]{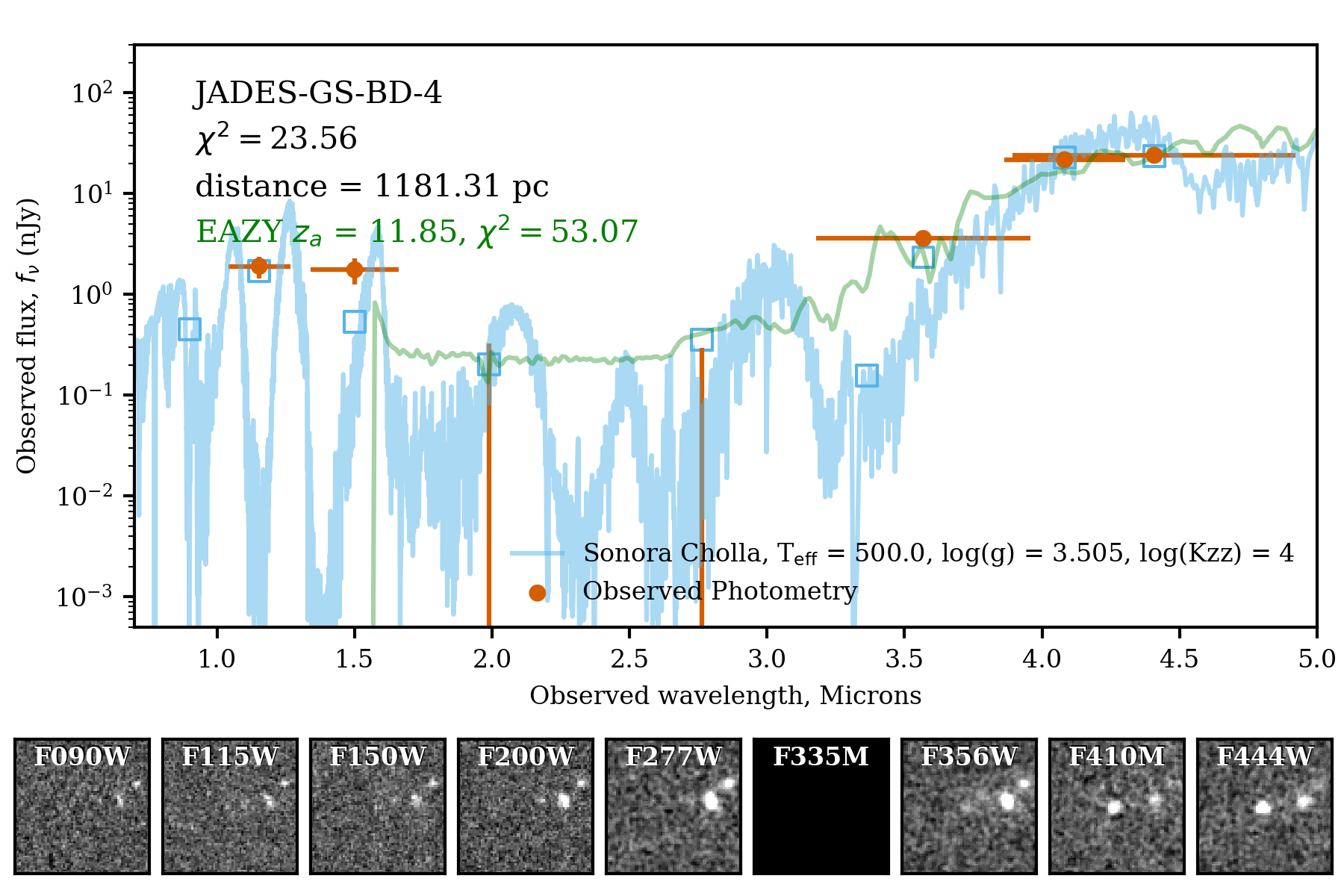}
  \includegraphics[width=0.49\textwidth]{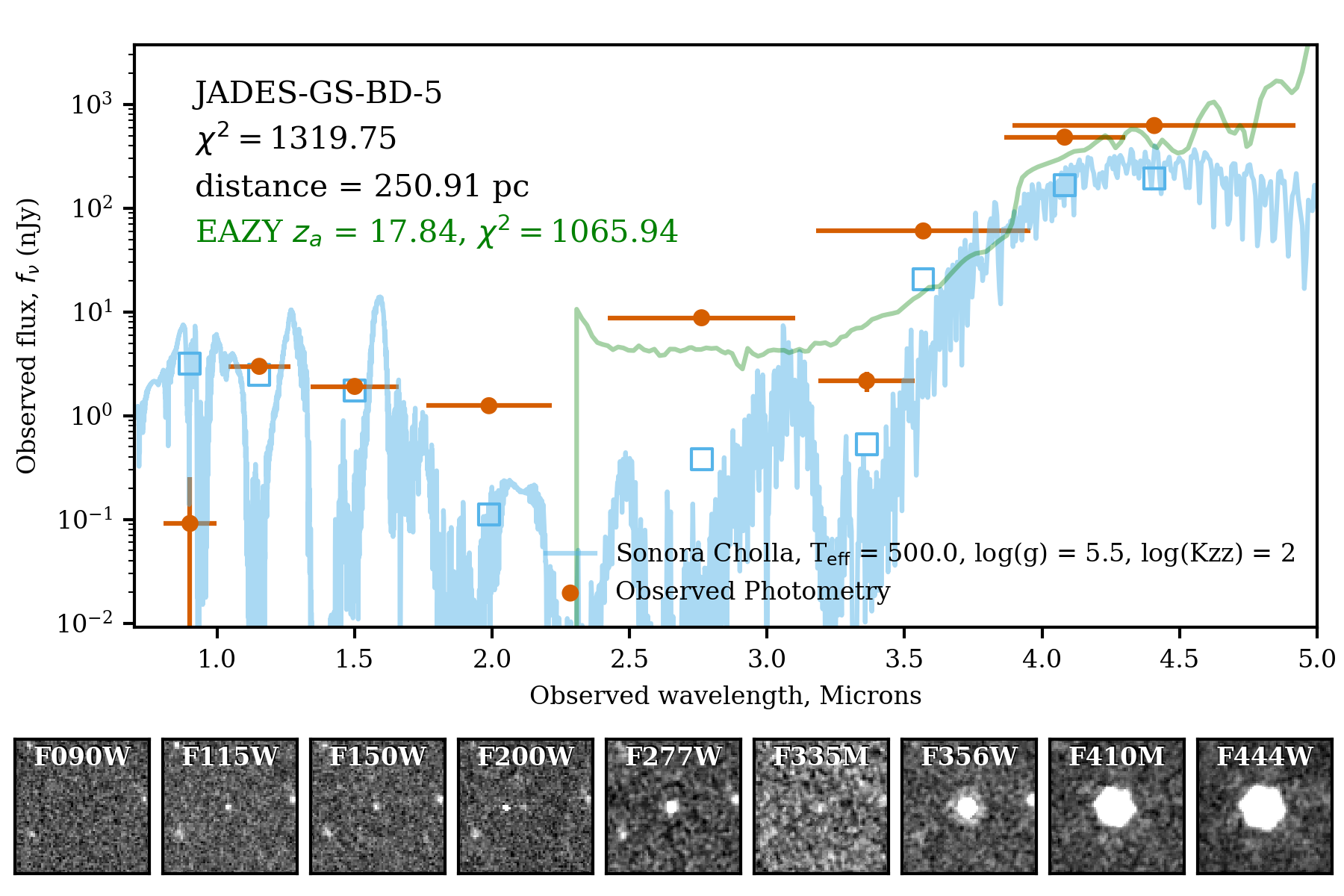}
  \includegraphics[width=0.49\textwidth]{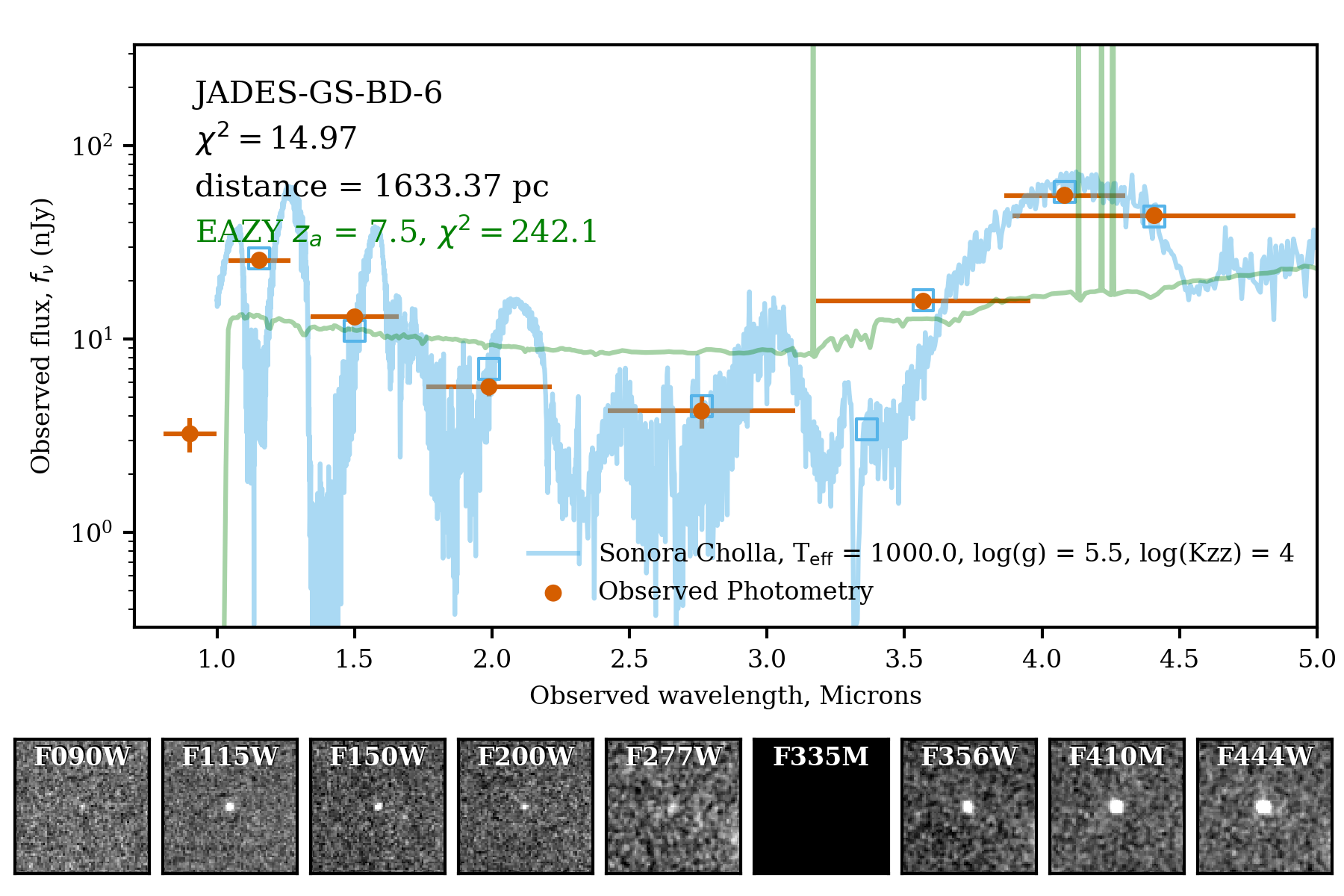}
  \caption{Brown dwarf candidate SEDs measured using {\tt forcepho} (red) with best-fits from the Sonora-Cholla substellar atmospheric models (blue) overplotted. We show the parameters for the model in the bottom right, and the $\chi^2$ of the fit and derived distance based on the model in the upper left. We plot the best-fitting {\tt EAZY} fit in green, and provide the photometric redshift derived from the fit and the resulting $\chi^2$. Beneath each SED plot are the individual JWST NIRCam thumbnails ($2^{\prime\prime} \times 2^{\prime\prime}$) for each source. This is continued in Figure \ref{fig:BD_SEDs_2} and \ref{fig:BD_SEDs_3}. \label{fig:BD_SEDs_1}}
\end{figure*}

\begin{figure*}
  \centering
  \includegraphics[width=0.49\textwidth]{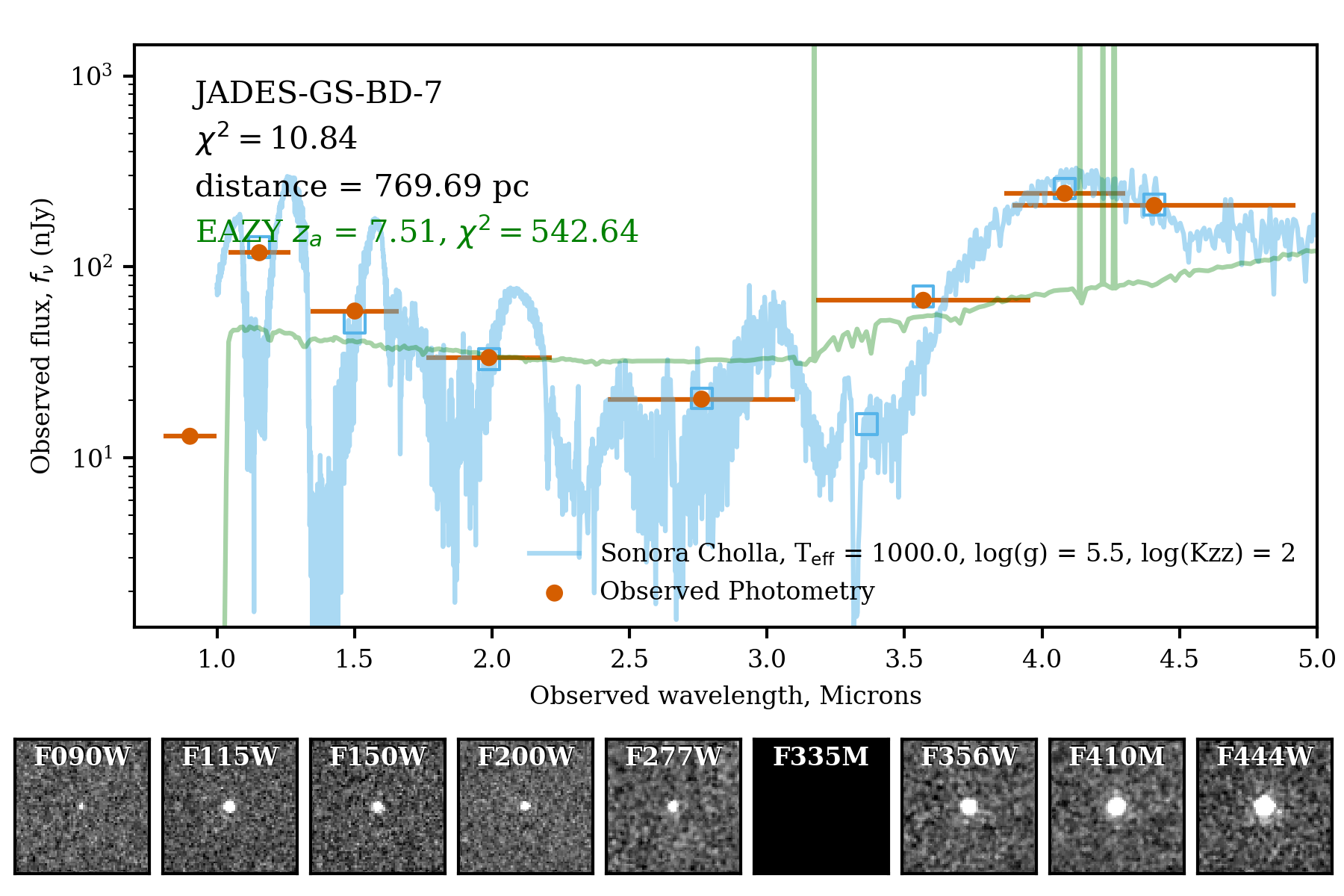}
  \includegraphics[width=0.49\textwidth]{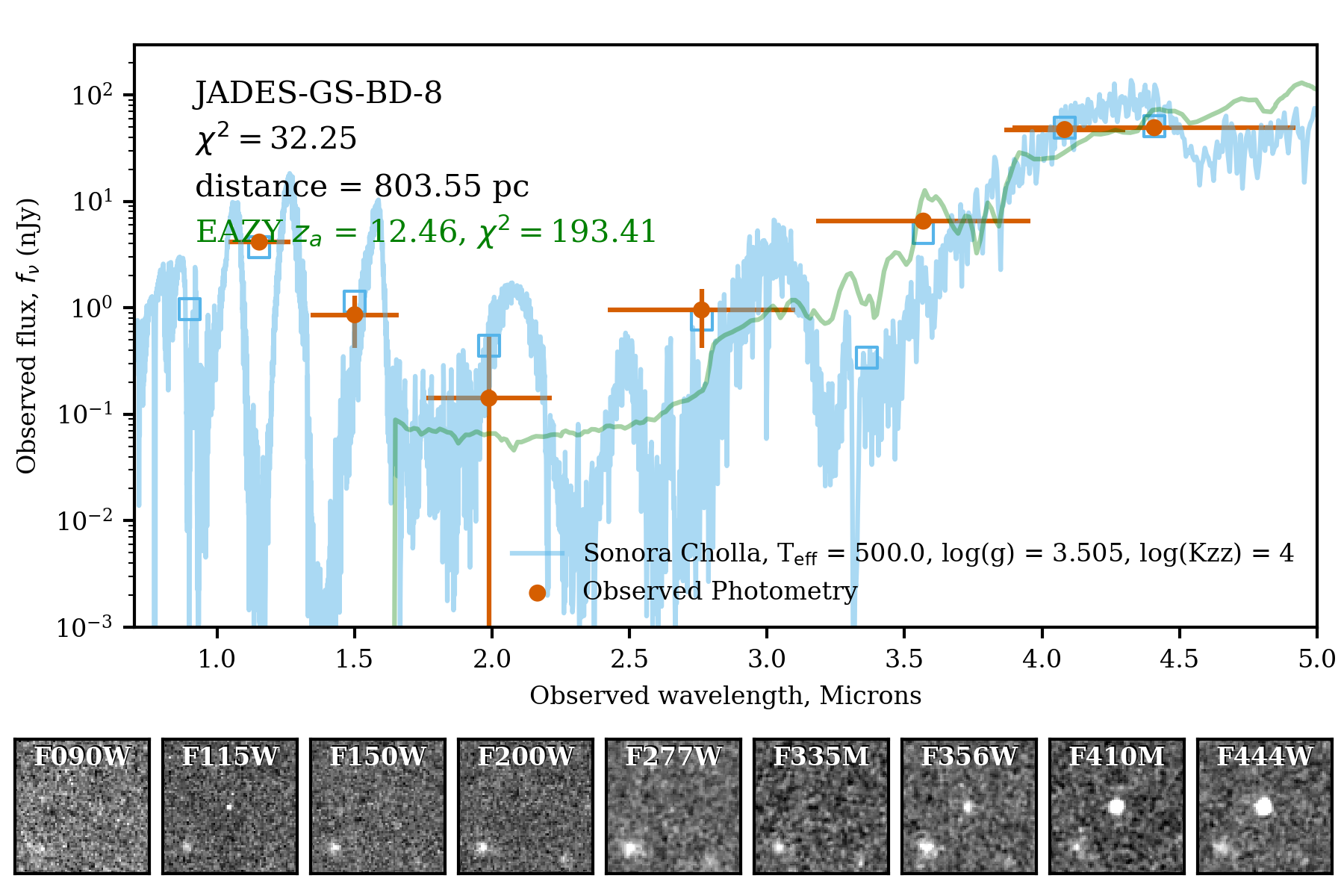}
  \includegraphics[width=0.49\textwidth]{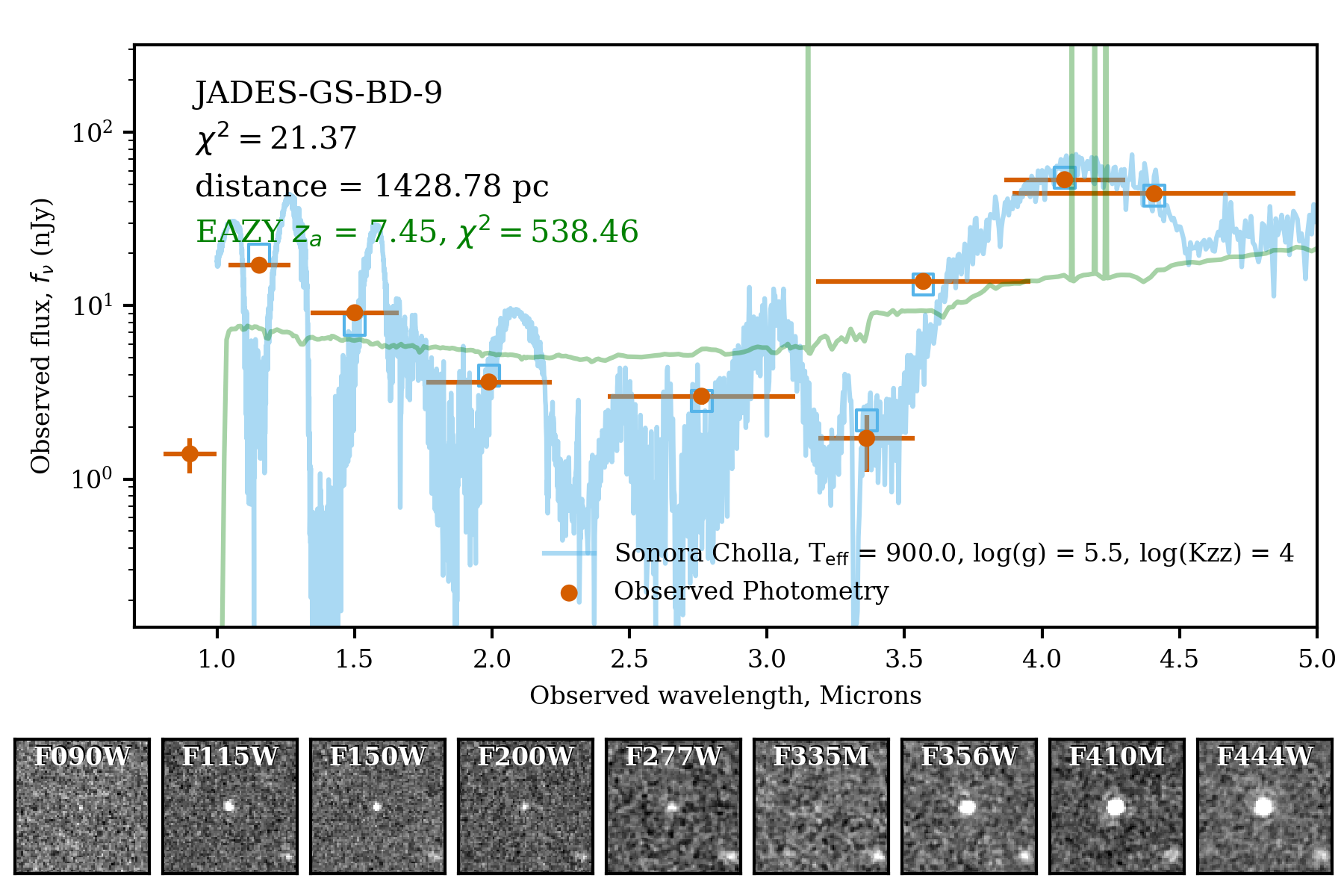}
  \includegraphics[width=0.49\textwidth]{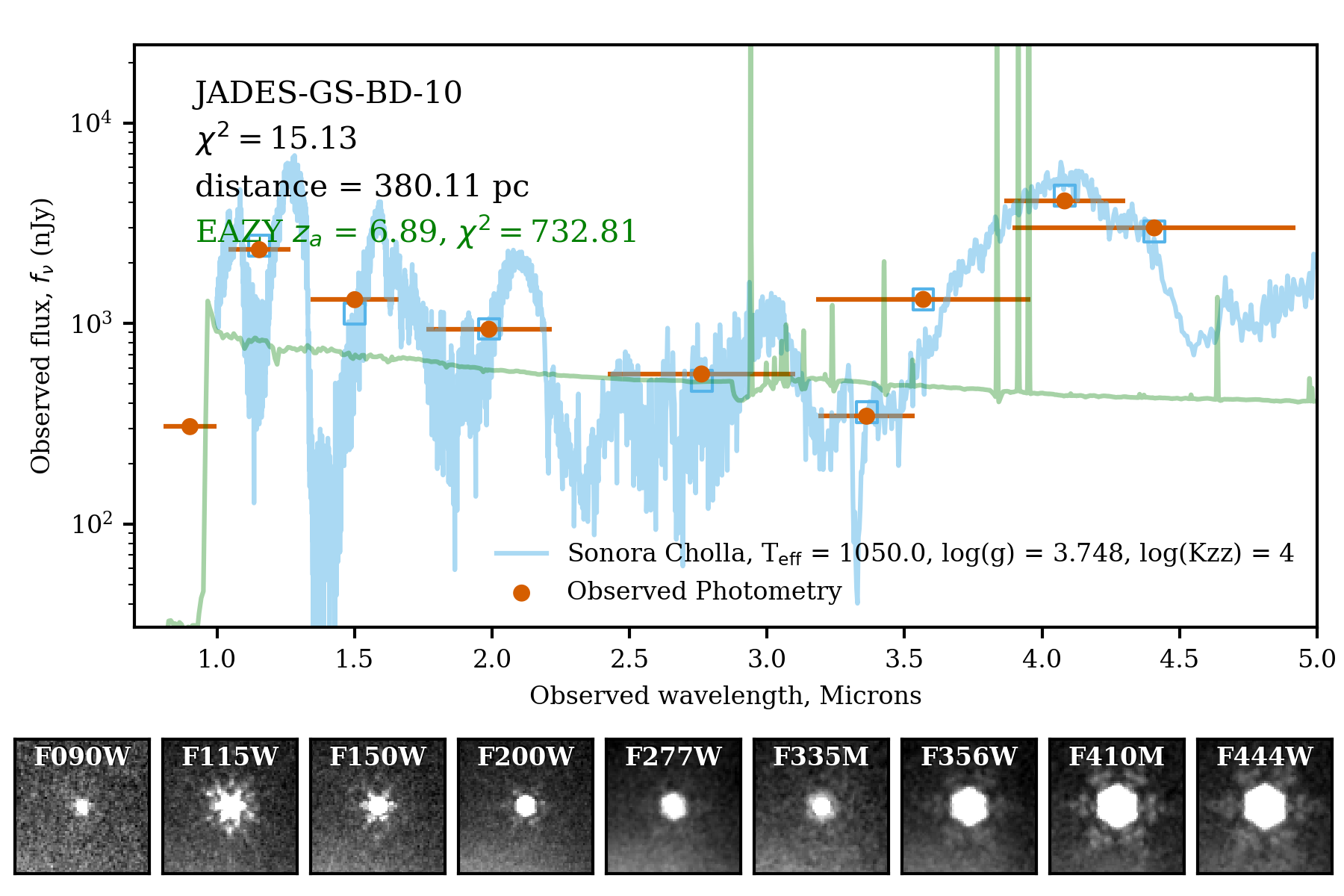}
  \includegraphics[width=0.49\textwidth]{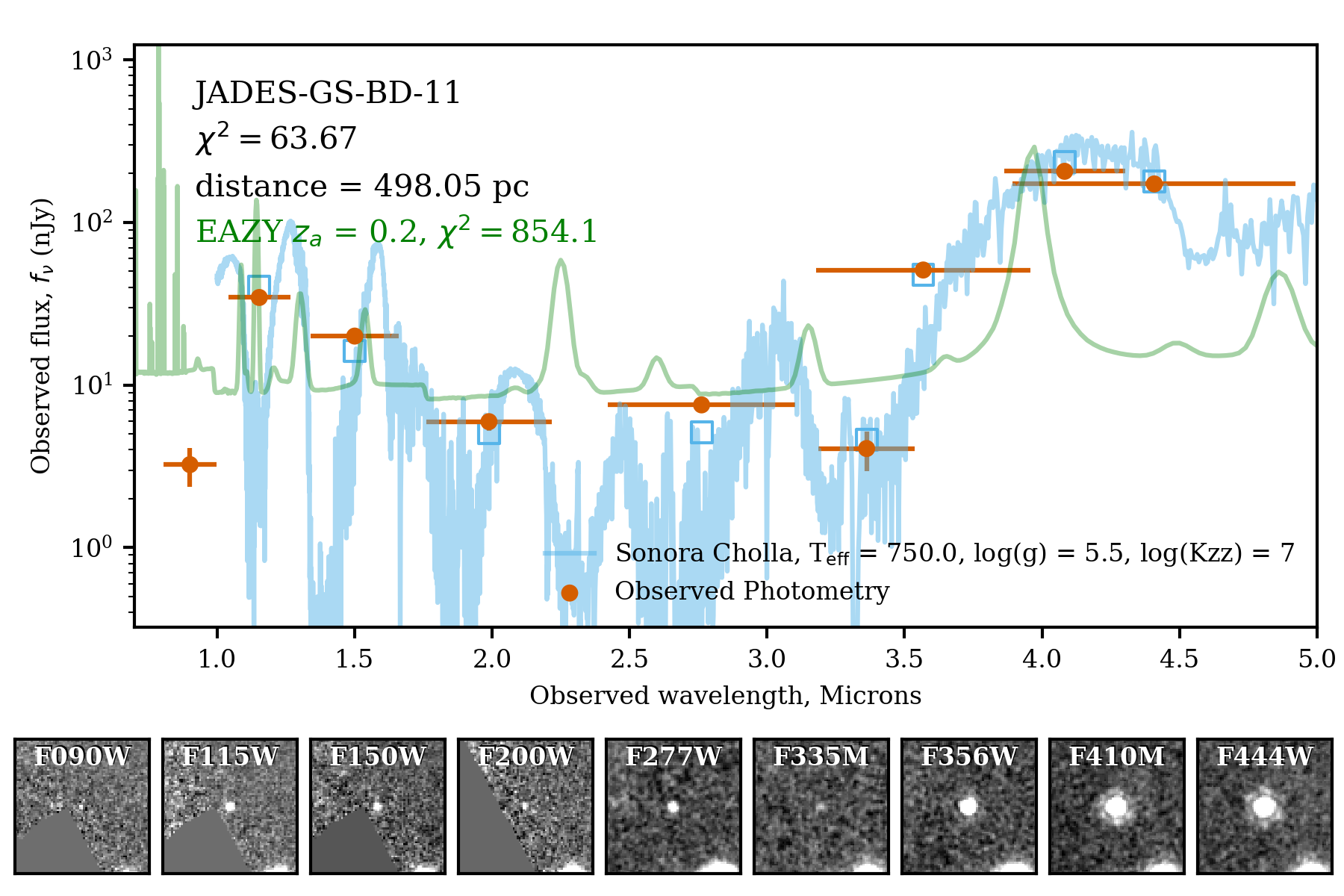}
  \includegraphics[width=0.49\textwidth]{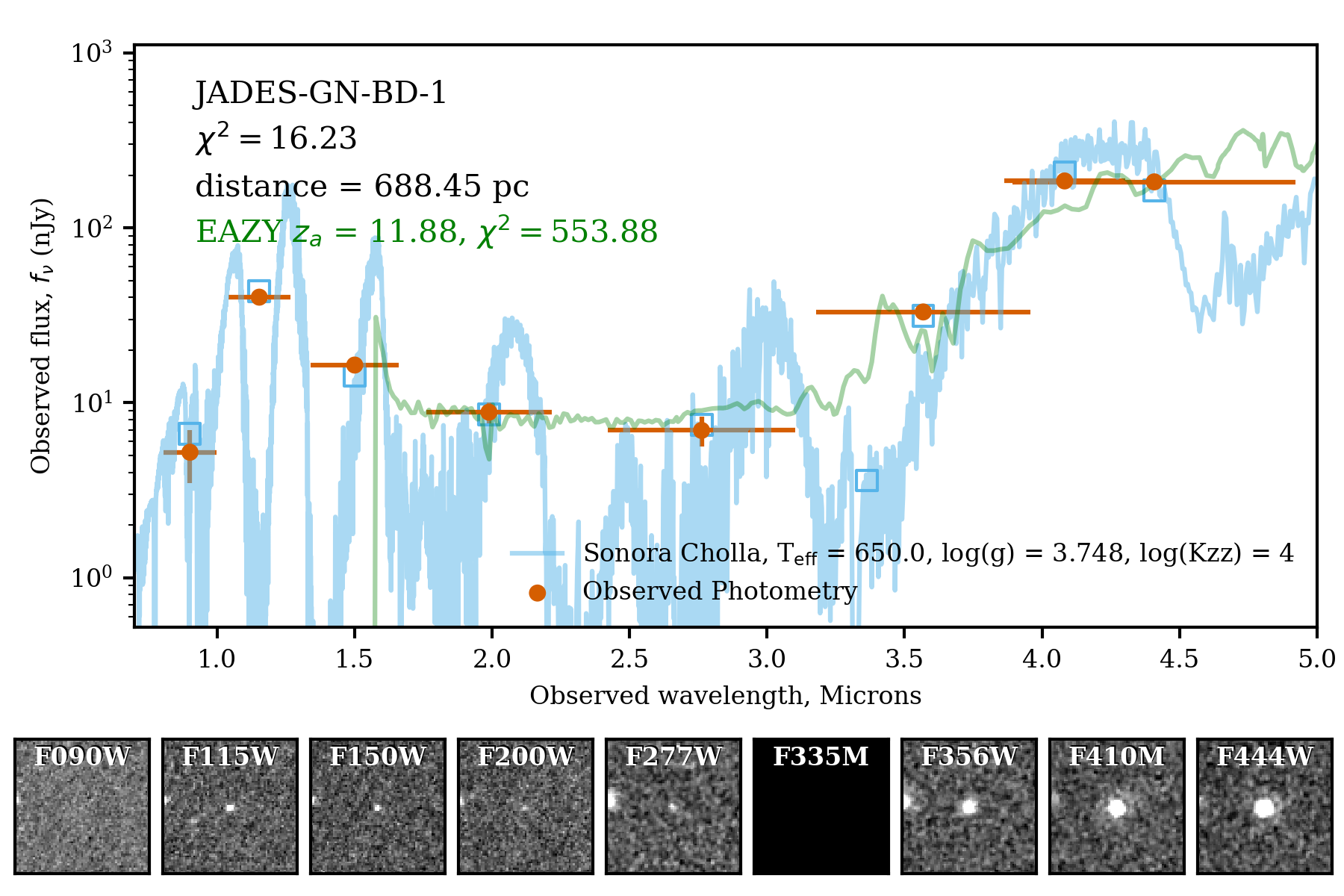}
  \includegraphics[width=0.49\textwidth]{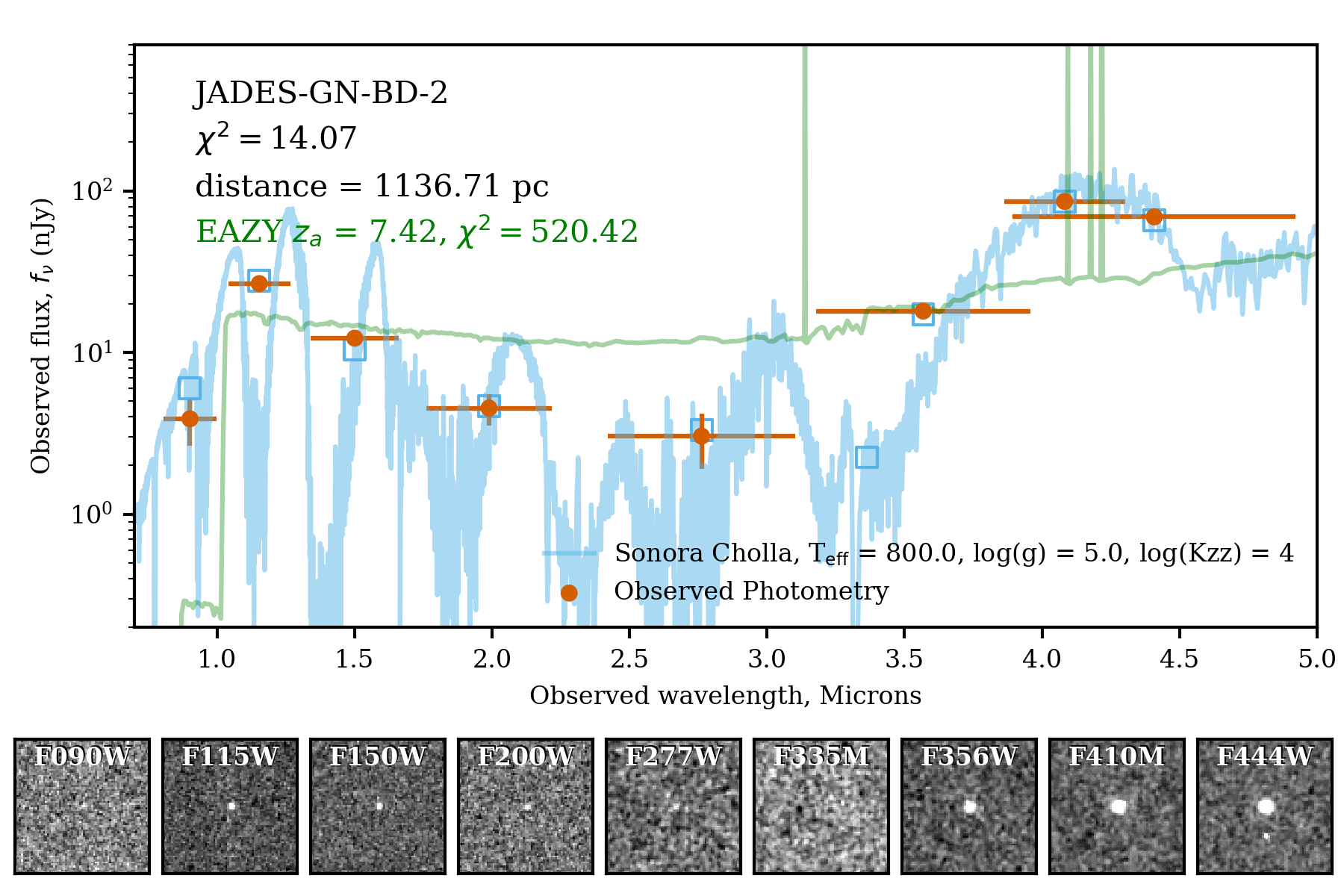}
  \includegraphics[width=0.49\textwidth]{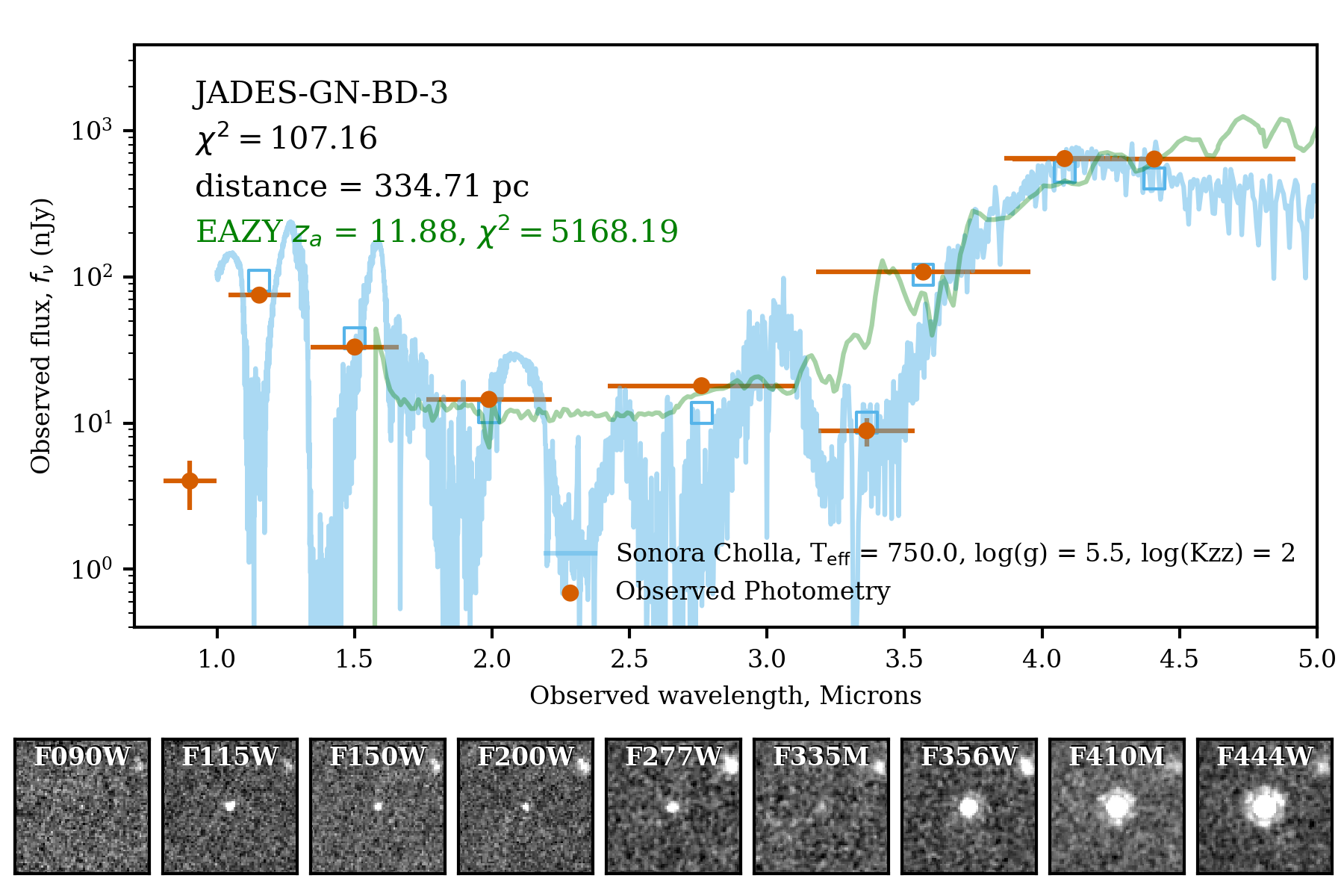}  
  \caption{Brown Dwarf Candidate SEDs (continued from Figure \ref{fig:BD_SEDs_1}).\label{fig:BD_SEDs_2}}
\end{figure*}

\begin{figure*}
  \centering
  \includegraphics[width=0.49\textwidth]{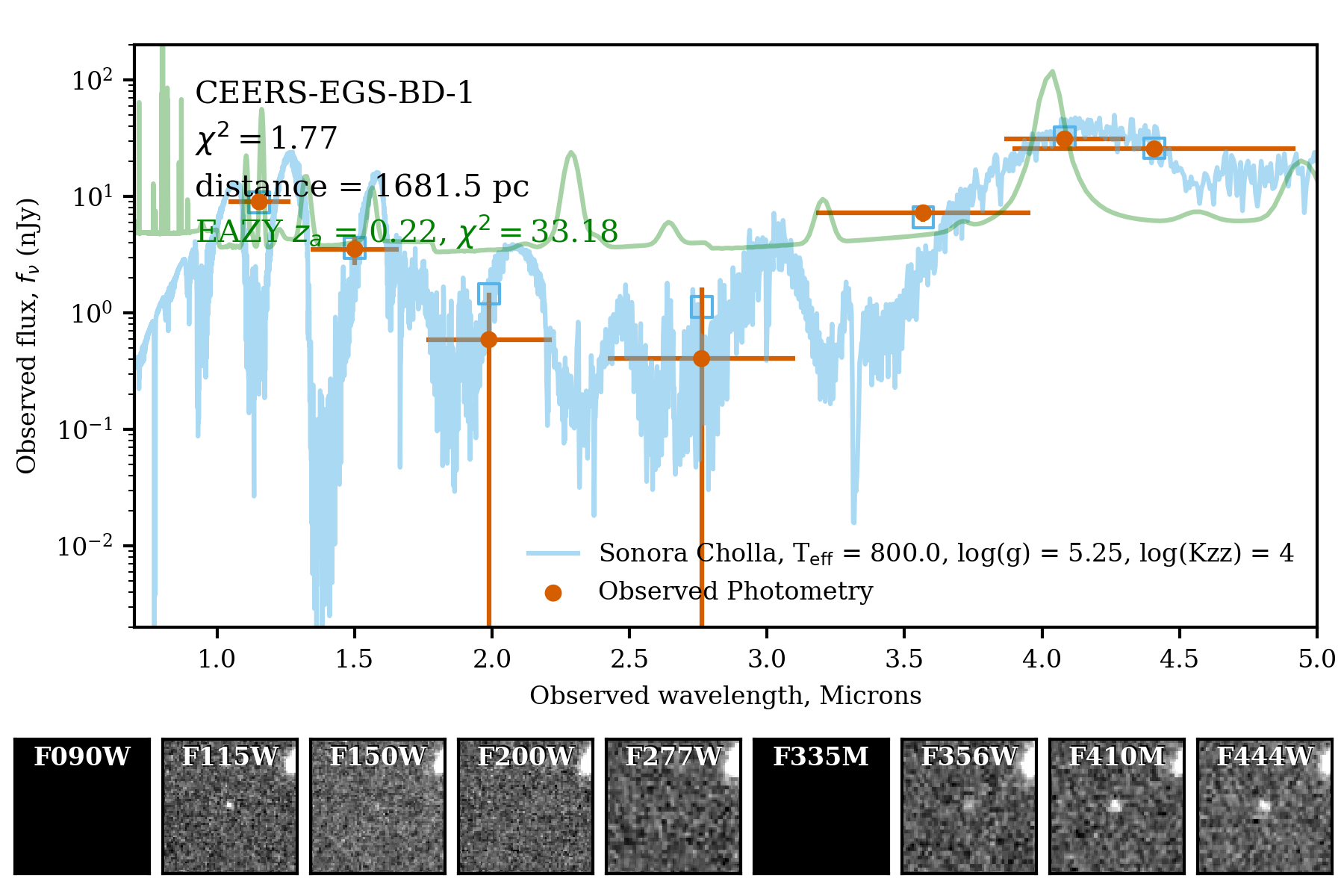}
  \includegraphics[width=0.49\textwidth]{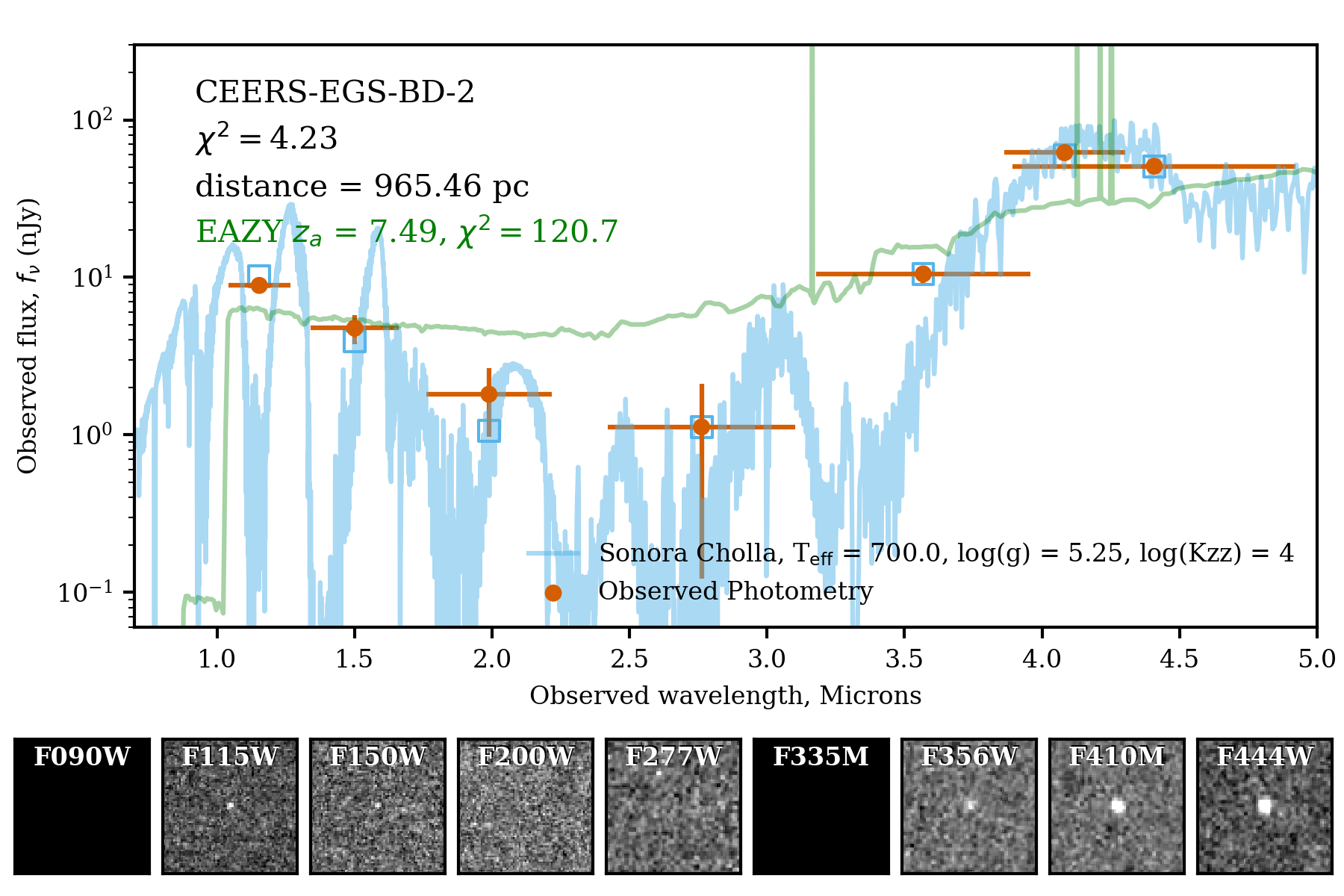}
  \includegraphics[width=0.49\textwidth]{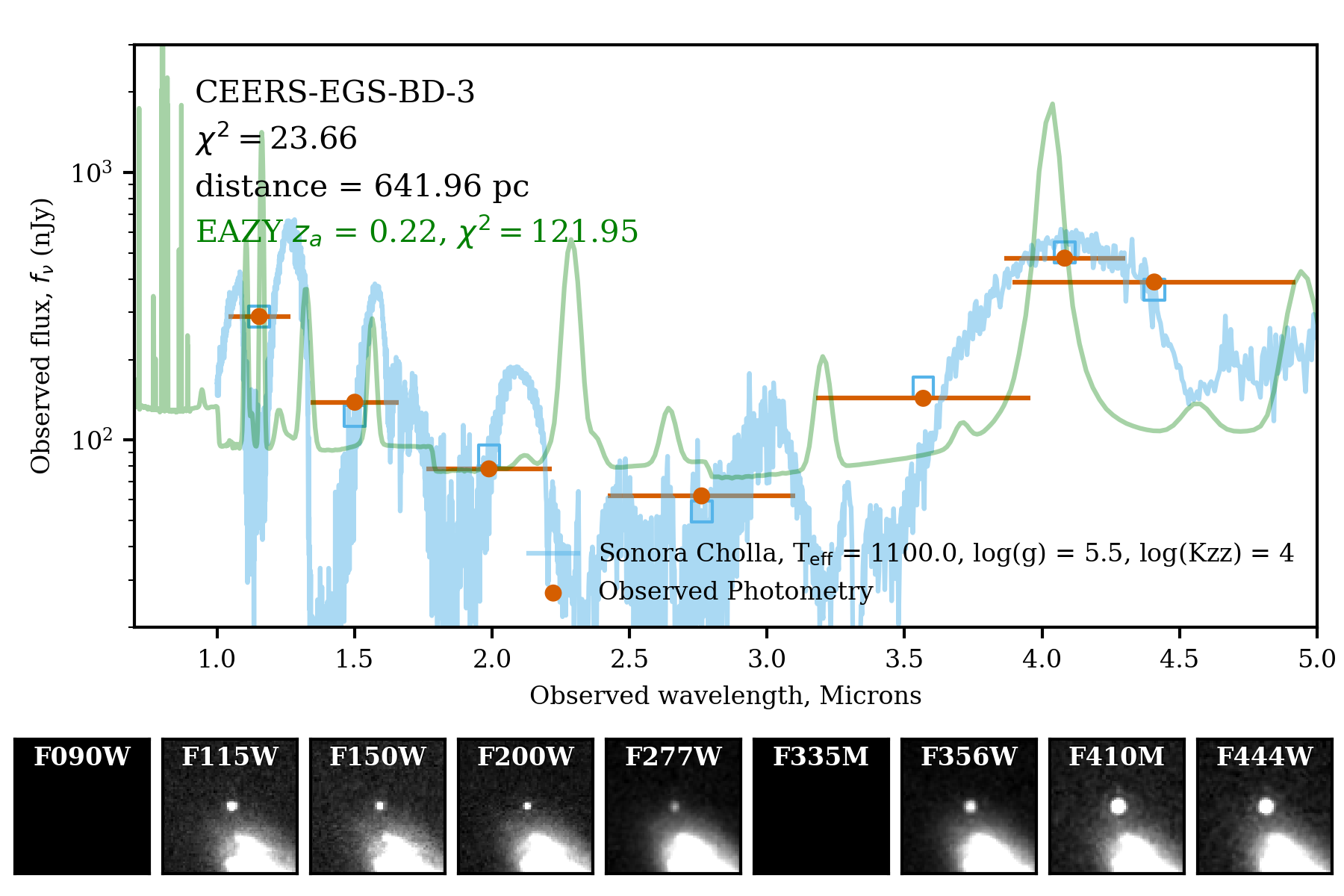}
  \includegraphics[width=0.49\textwidth]{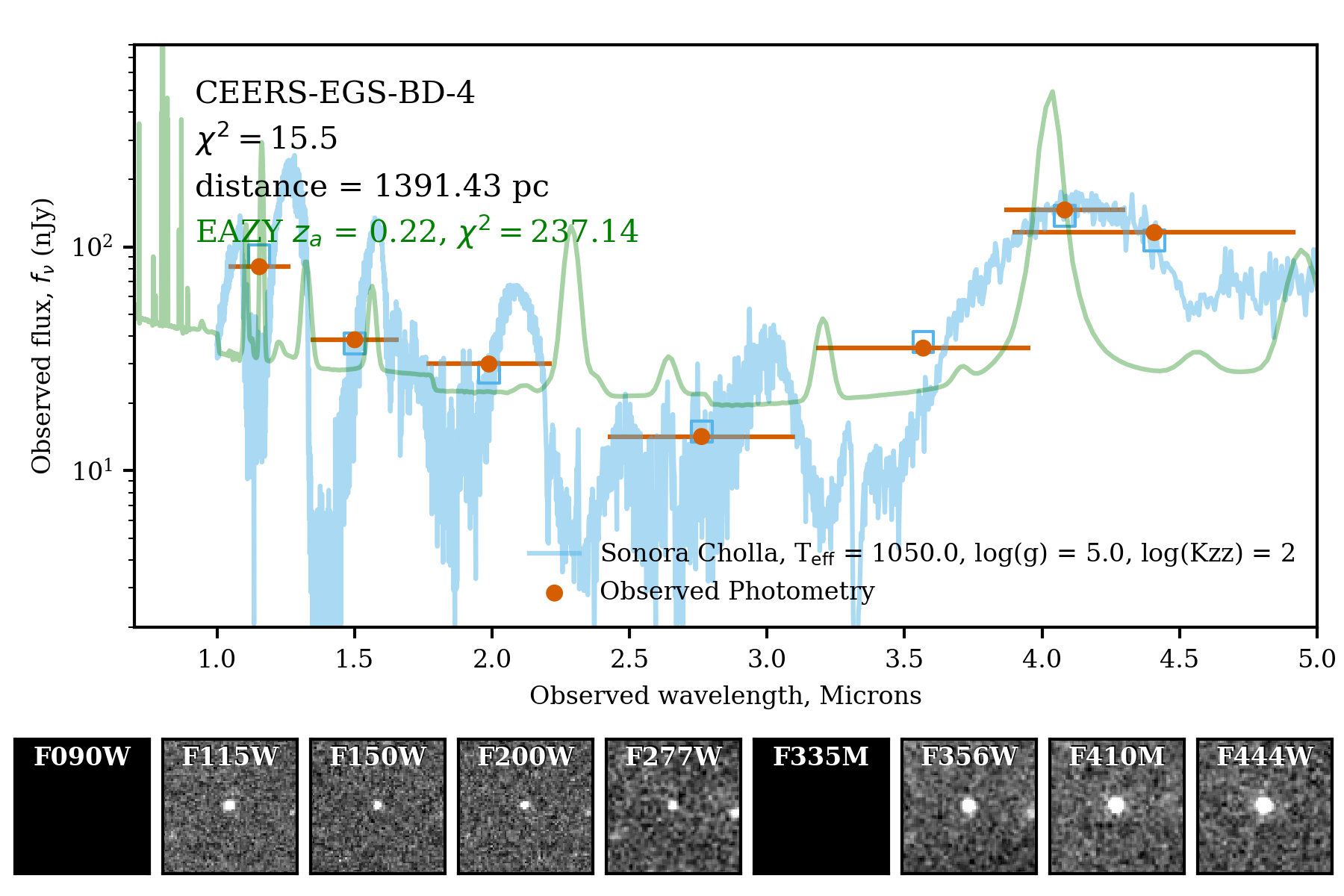}
  \includegraphics[width=0.49\textwidth]{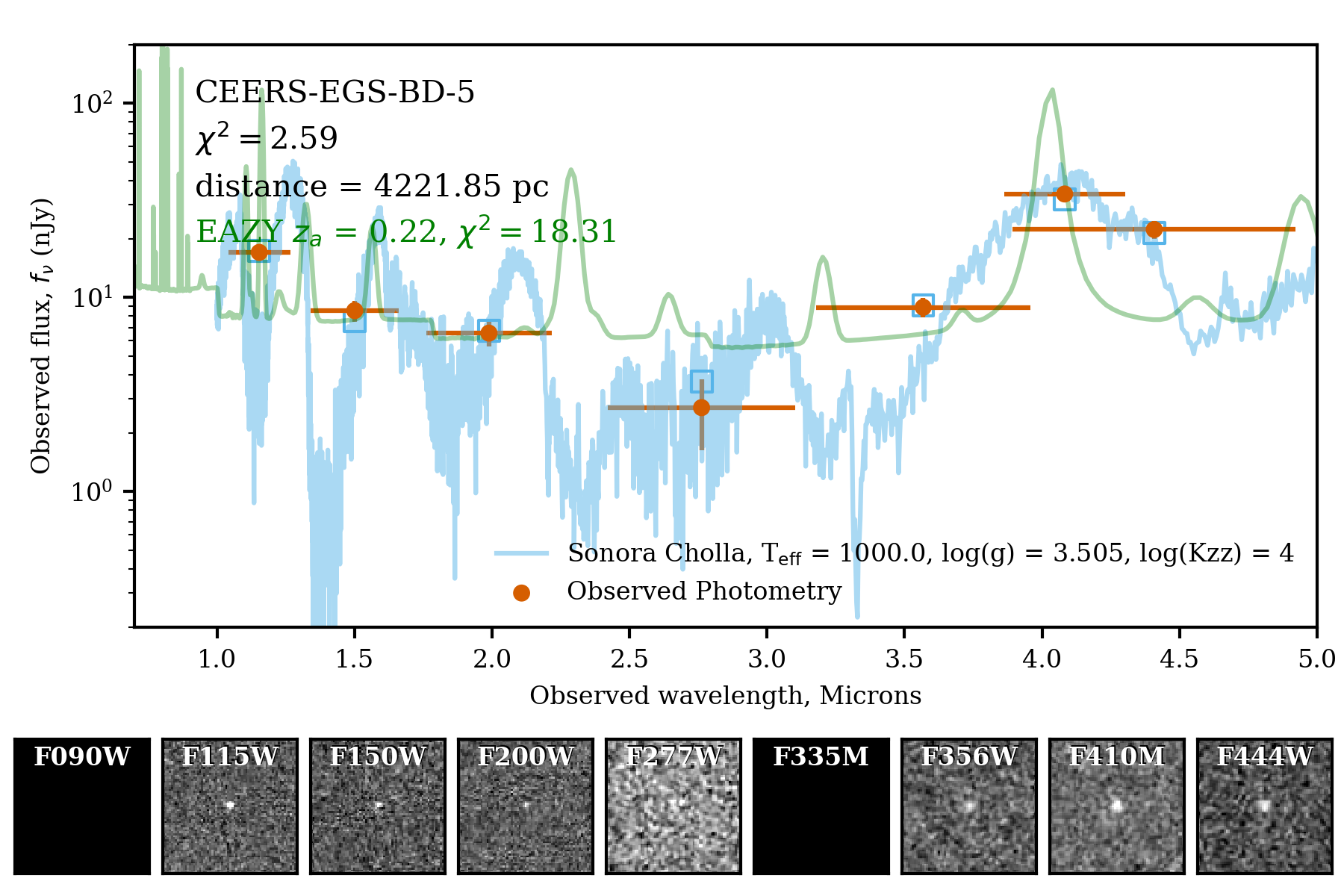}
  \includegraphics[width=0.49\textwidth]{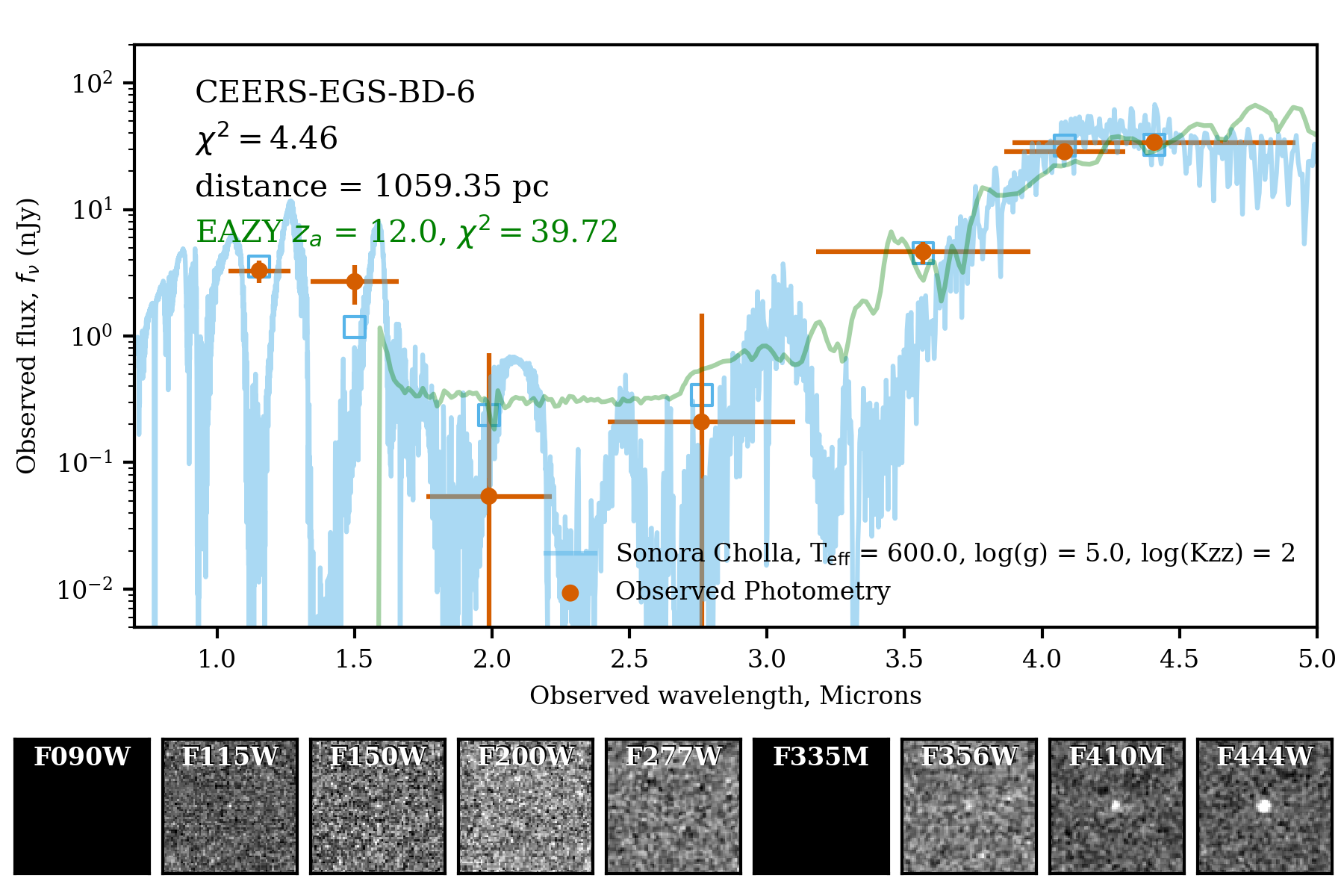}
  \includegraphics[width=0.49\textwidth]{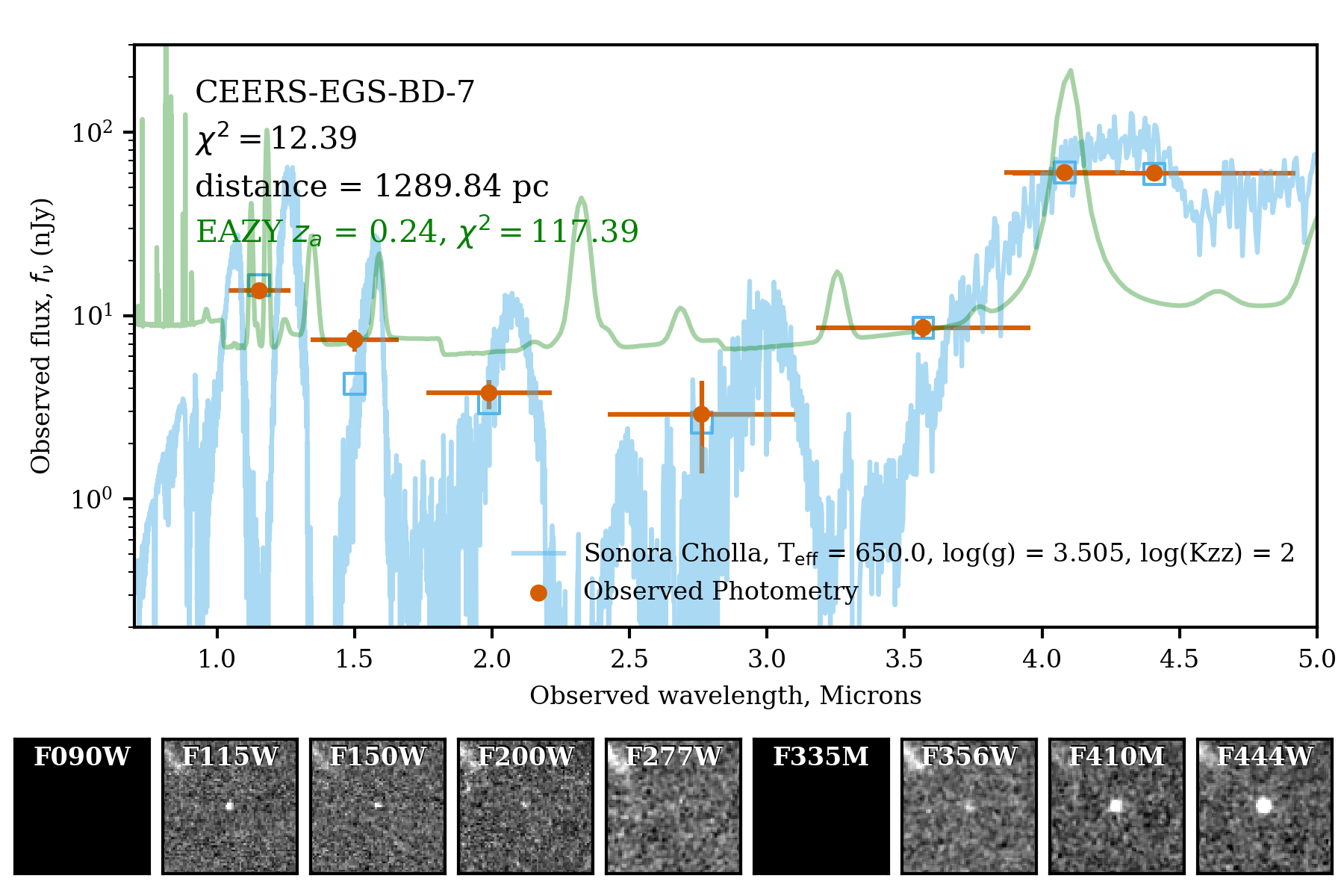}
  \caption{Brown Dwarf Candidate SEDs (continued from Figure \ref{fig:BD_SEDs_2}).\label{fig:BD_SEDs_3}}
\end{figure*}

We present the full list of JADES and CEERS brown dwarf candidates in Table \ref{tab:bd_fluxes}, with the {\tt forcepho} fluxes and effective radii. Our naming convention starts with the survey name (JADES or CEERS), the region of the sky (GOODS-S, GOODS-N, or EGS), ``BD'' indicating brown dwarf candidate, and a unique integer based on the RA of the source. We plot the SEDs for the selected brown dwarf candidates in Figures \ref{fig:BD_SEDs_1}, \ref{fig:BD_SEDs_2}, and \ref{fig:BD_SEDs_3} and indicate T$_{\mathrm{eff}}$, log $g$, log $K_{zz}$, and  minimum $\chi^2$ of the best-fitting Sonora Cholla model. In each panel, we also provide the minimum $\chi^2$ derived from the {\tt EAZY} galaxy fits to the source, the best-fitting photometric redshift, and show the galaxy template fit for each object. 

To estimate the distances to the BD candidates based on the fits, we followed the procedure detailed in \citet{wang2023}. The best-fit normalization of the Sonora Cholla models to the observed photometry is equal to $(R/D)^2$, where $R$ is the radius of the model source and $D$ is the distance to the object. We used the ``Sonora Bobcat'' evolution models \citep{marley2021} to find the brown dwarf radii corresponding to the best-fit T$_{\mathrm{eff}}$ and log $g$ values for both solar metallicity ([M/H] $= 0$) and subsolar metallicity ([M/H] $= -0.5$) models. Evolution models use energy loss rate estimates to explore how brown dwarf radii and luminosity change over time. Because the \citep{marley2021} Sonora Bobcat evolution models to do not include radii for brown dwarfs with log $g$ = 5.5, we extrapolate the models to estimate those values. For the ATMO 2020 fits, we followed the same procedure and estimated the radius of each source from the provided evolutionary models, which allowed us to calculate distances. 

As part of the Sonora Bobcat model data release, the authors provide observed photometry at 10 pc for different values of T$_{\mathrm{eff}}$, log $g$, and radius. To demonstrate the uncertainty in distance estimates to these sources, we additionally fit our candidate photometry directly to the Sonora Bobcat model photometry, and we derive a third distance estimate from the scaling to the observed photometry for our candidates. 

In Table \ref{tab:bd_derived_properties}, we provide the best-fitting parameters, the derived substellar radii, and the distances for our candidates for the Sonora Cholla and ATMO 2020 fits. We also provide the distances and parameters from fitting directly to the Sonora Bobcat model photometry. Using the subsolar Sonora Bobcat models results in distance measurements that are only 0.6\% smaller. We do see a more significant difference when comparing to the distances derived directly from the Sonora Bobcat evolutionary models, which are, on average, 11\% larger than what we estimate using the Sonora Cholla models. Fits with the ATMO 2020 models result in distances that are 17\% larger than those derived using the solar metallicity Sonora Cholla models. At these best-fit effective temperatures, adopting the Sonora Cholla fit values, fourteen of these these sources are T dwarf candidates (700K to 1300K), and the remaining seven are Y dwarf candidates ($<$700K). 

In addition, in Table \ref{tab:bd_fluxes}, we also provide the measured {\tt forcepho} half-light radii for these candidates. Each of the JADES and CEERS candidates in our sample has a half-light radius less than 0.005$^{\prime\prime}$, which strongly supports the argument that they are unresolved. In addition, the image thumbnails shown in Figures \ref{fig:BD_SEDs_1}, \ref{fig:BD_SEDs_2}, and \ref{fig:BD_SEDs_3} often have long wavelength morphologies with observable diffraction features, indicating the sources are unresolved. 

\subsection{Proper Motions} 

\begin{figure*}
  \centering
  \includegraphics[width=0.49\textwidth]{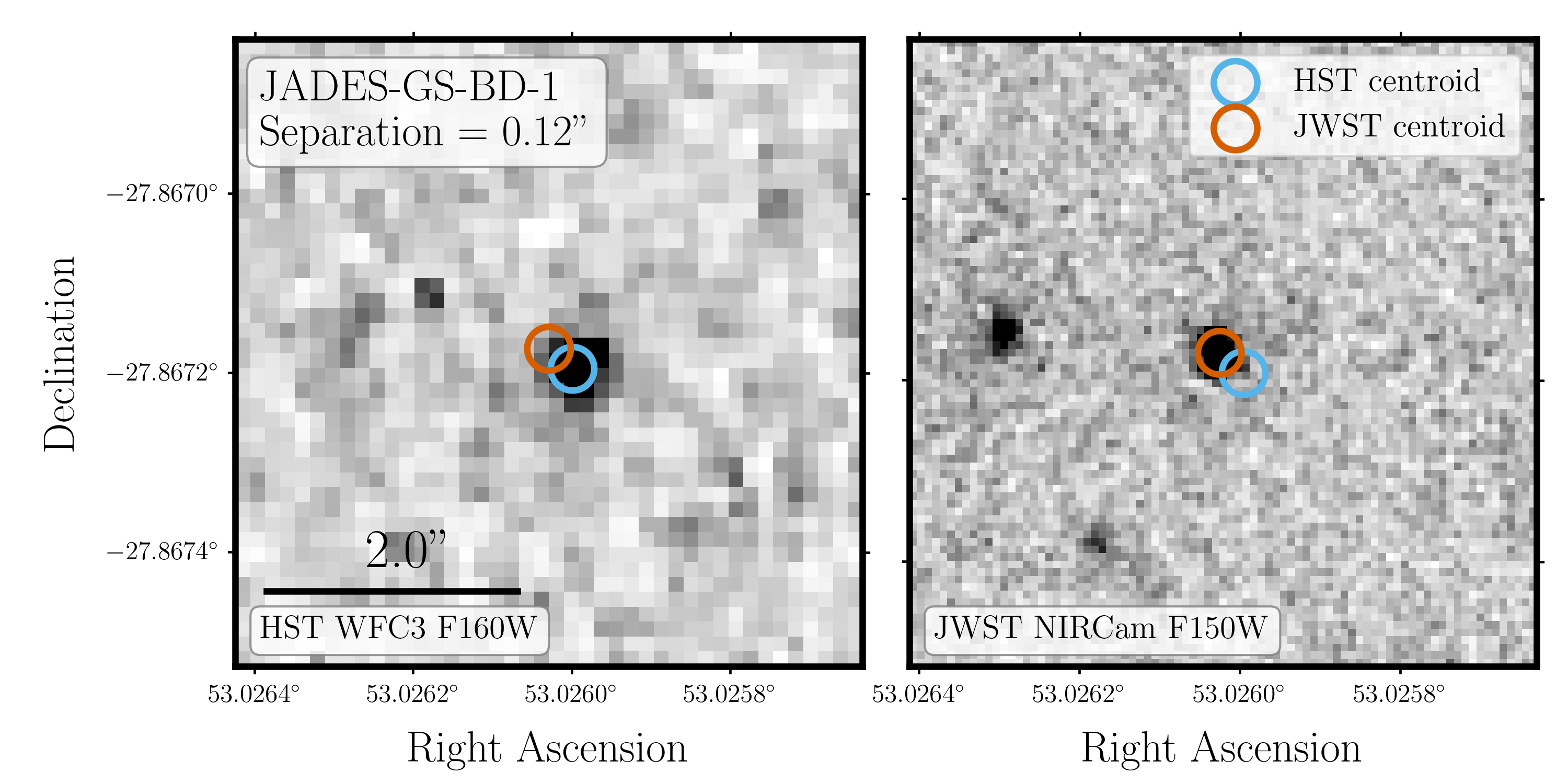}
  \includegraphics[width=0.49\textwidth]{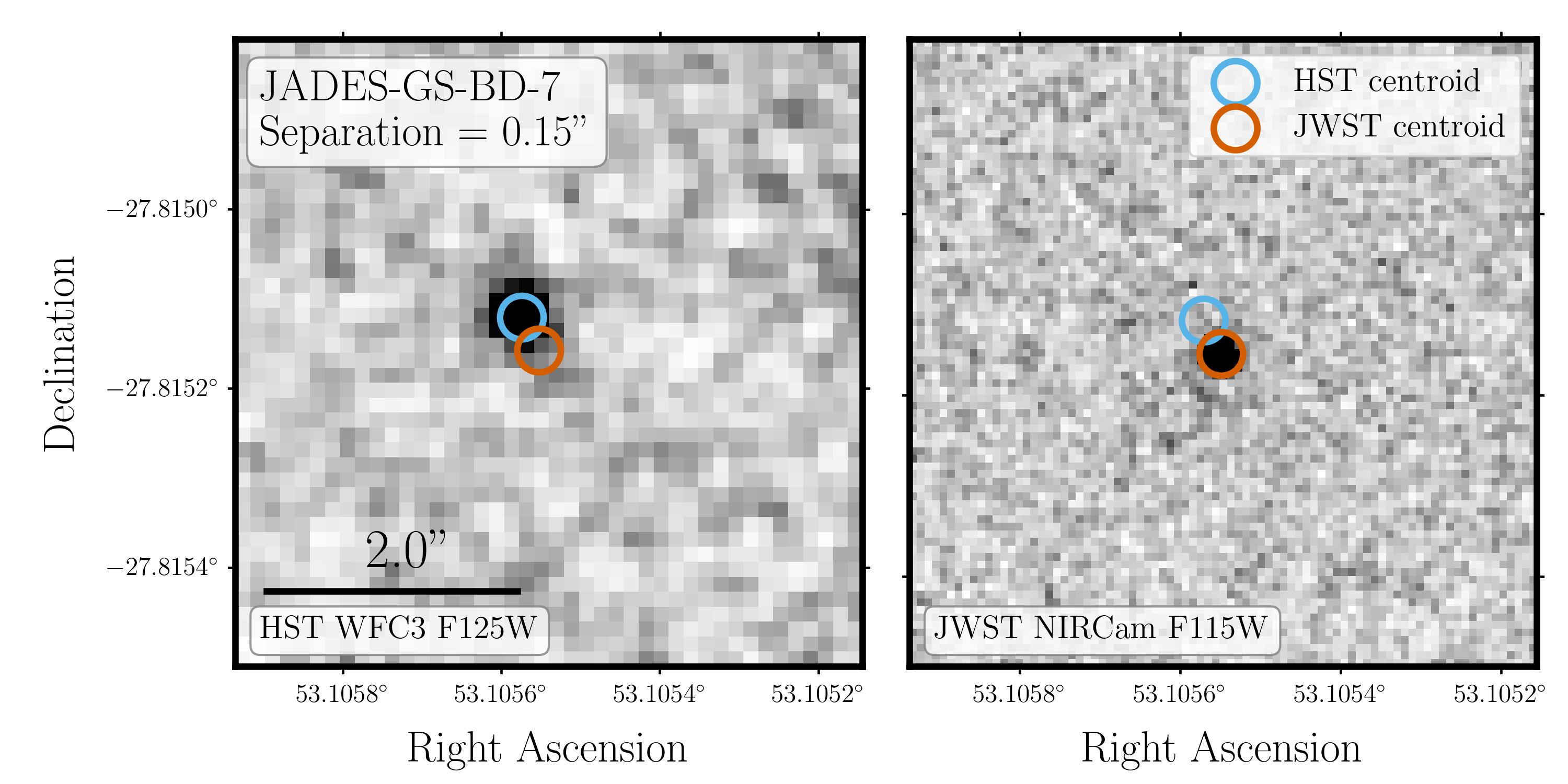}
  \includegraphics[width=0.49\textwidth]{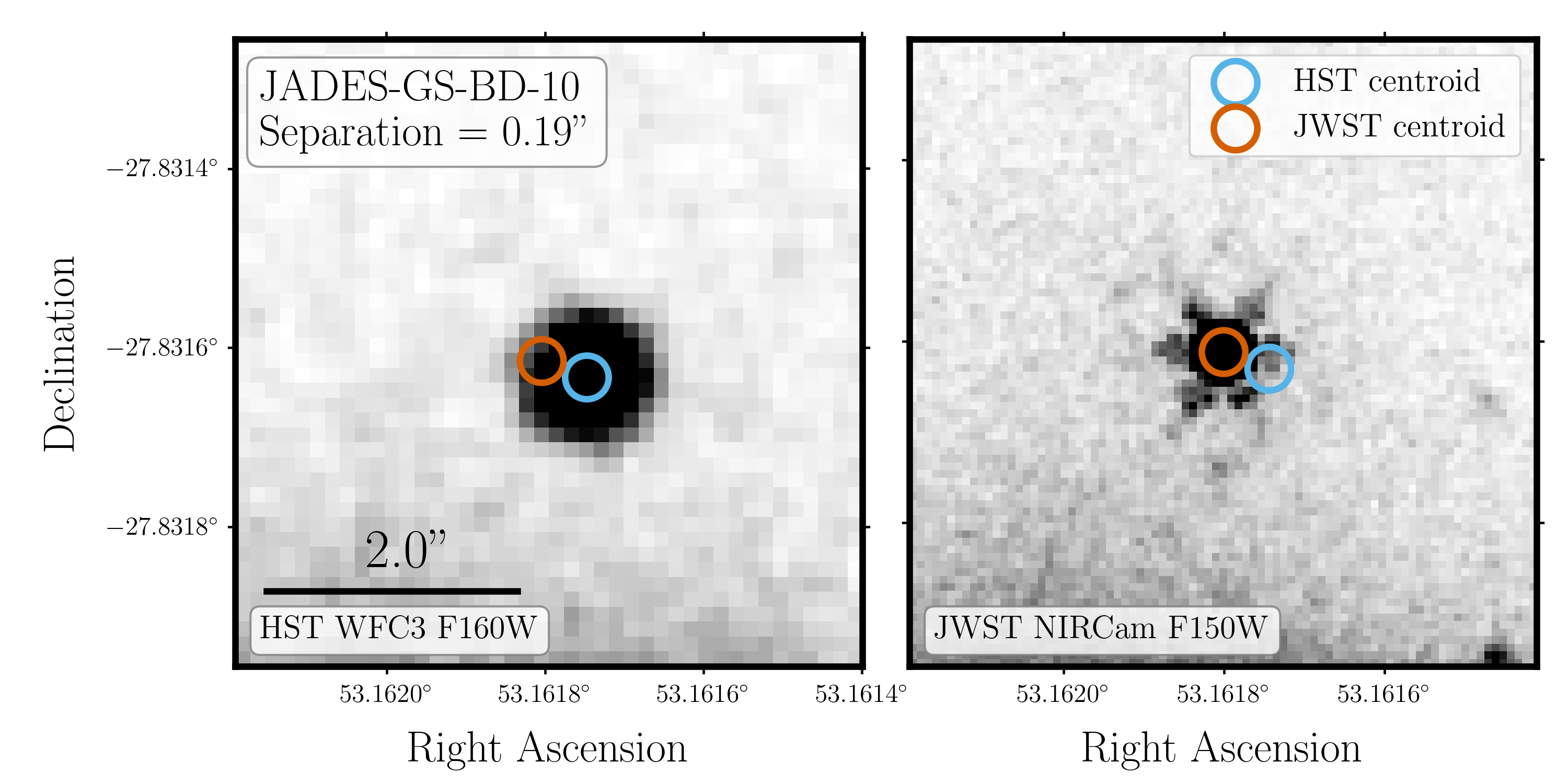}
  \includegraphics[width=0.49\textwidth]{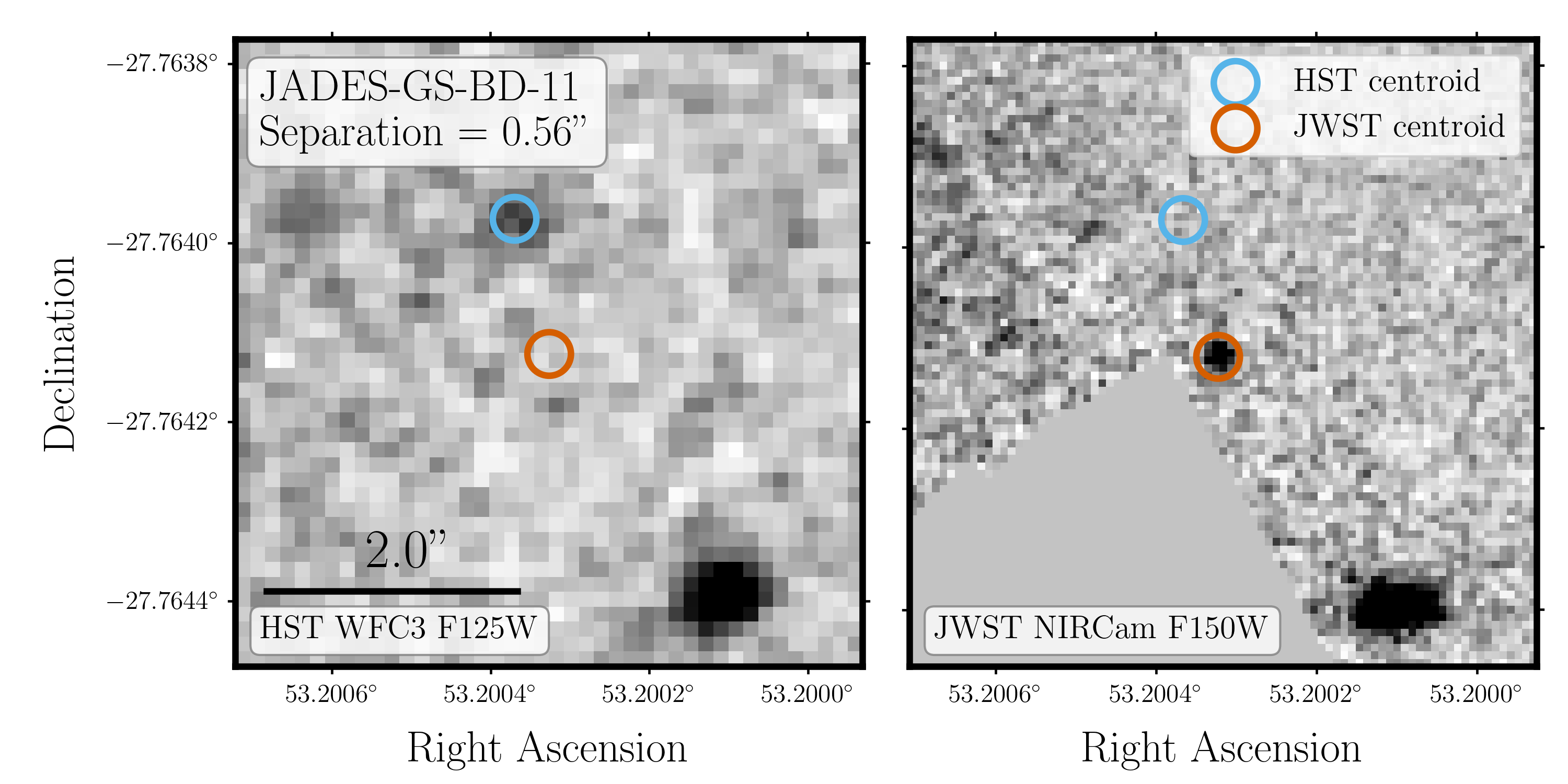}
  \includegraphics[width=0.49\textwidth]{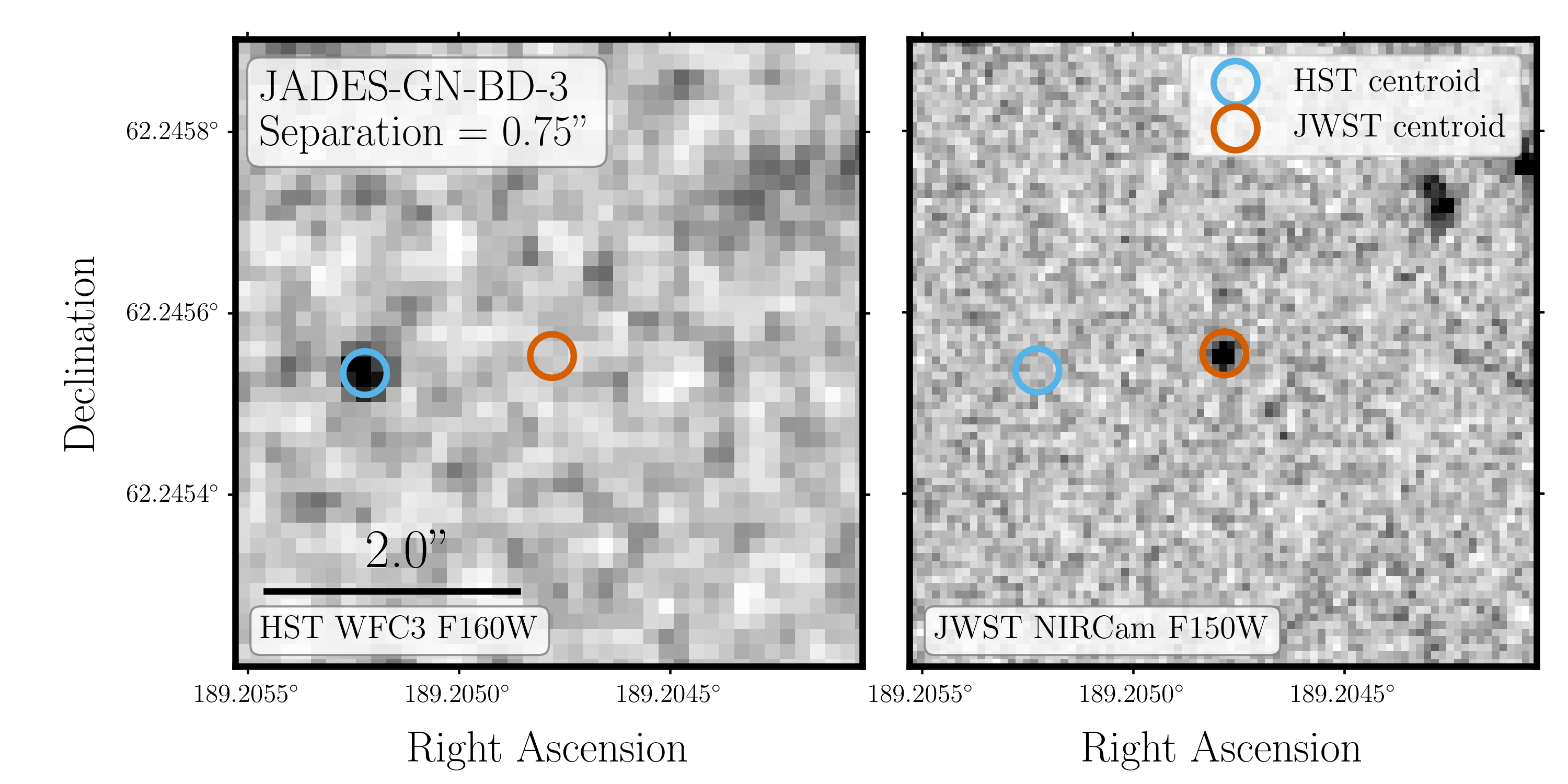}
  \includegraphics[width=0.49\textwidth]{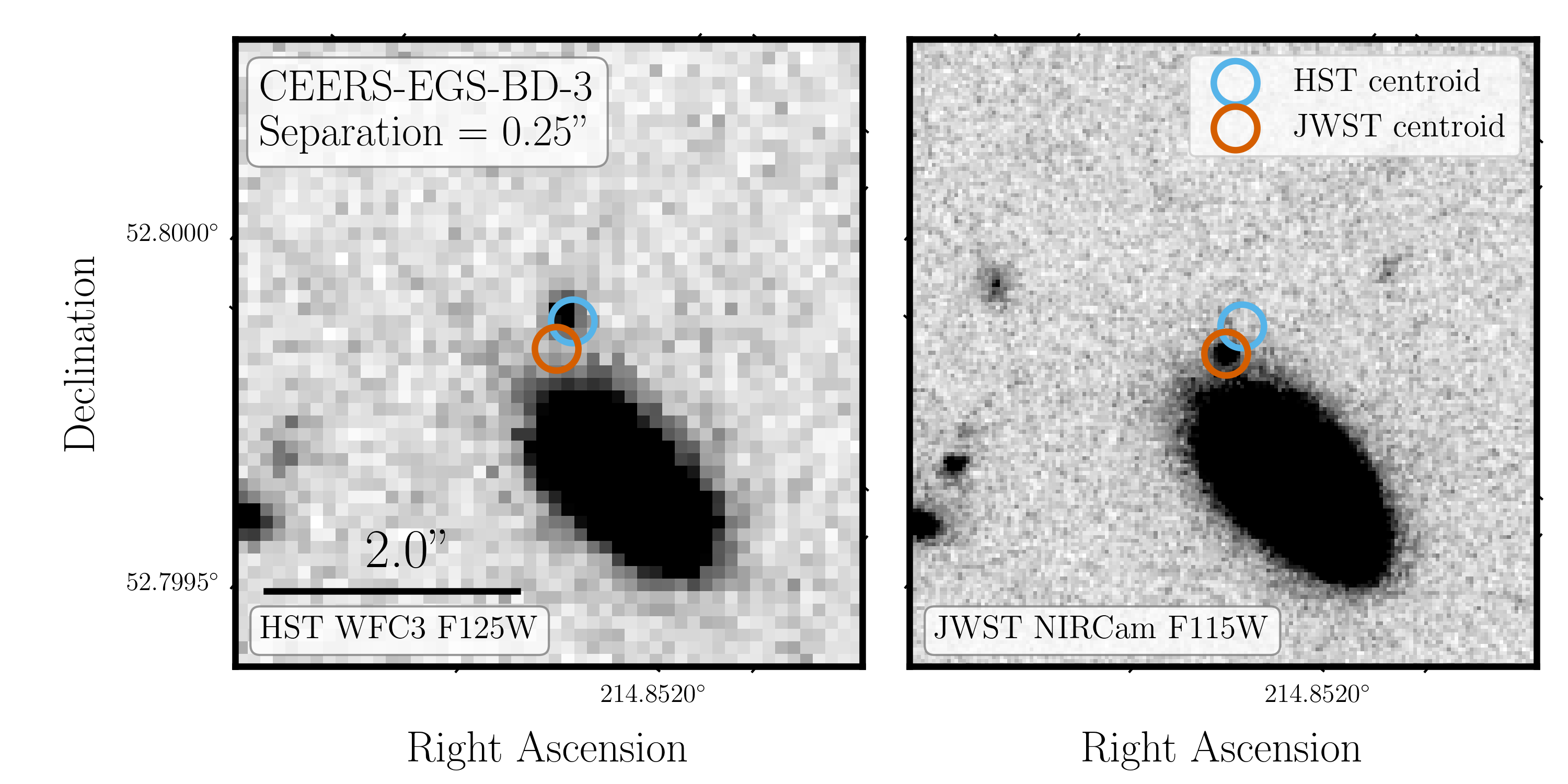}
  \includegraphics[width=0.49\textwidth]{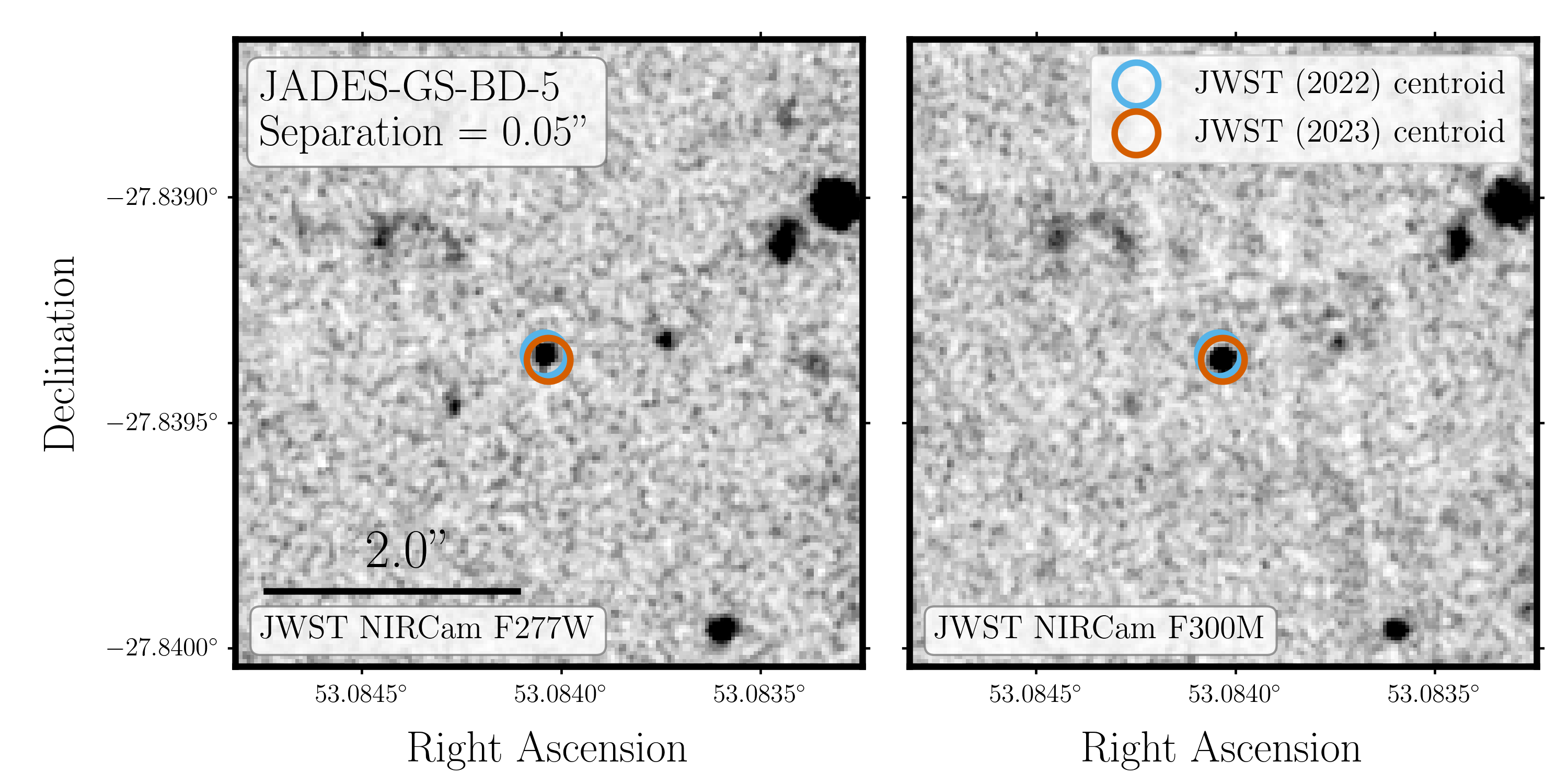}
  \includegraphics[width=0.49\textwidth]{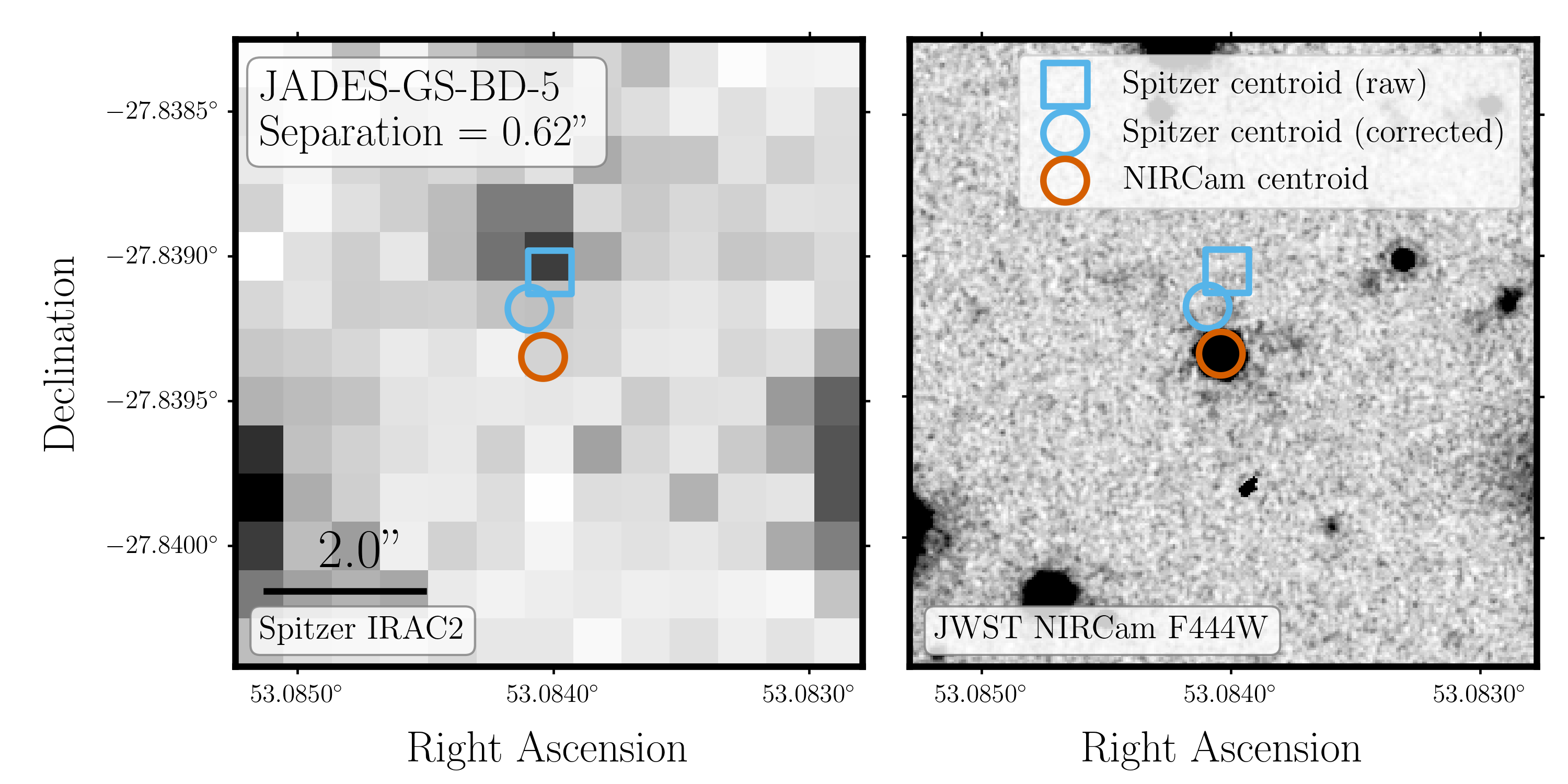}
  \caption{Proper motion measurements between the HST WFC3 and JWST NIRCam or between two epochs of JWST NIRCam observations. We plot seven sources, and for each source we plot the first epoch of observations in the left panel, and the second epoch in the right panel, with the filter being used shown in the bottom left. We also plot a scale bar in the left panels, above the filter name. In each panel, the centroid measured for the source from the HST, Spitzer, or JWST observations is shown with a blue circle, and the centroid for the source from the JWST observations is shown in a red circle. The offset, in arcseconds, is provided. For JADES-GS-BD-5, we show two independent measurements of the proper motion. The bottom left pair of panels show an offset between JADES data taken in late 2022 and observations as part of the JADES Origins Field \citep{eisenstein2023b} in late 2023. The bottom right pair of panels show a detected offset between Spitzer IRAC2 observations and the JADES F444W observation. Because the Spitzer IRAC2 image has a different WCS than the JADES NIRCam mosaic, we measure the raw centroid and label it with a square symbol, and the centroid corrected to the JADES WCS is plotted with a circle, along with the corresponding separation. \label{fig:BD_Proper_Motions}}
\end{figure*}

The JADES regions, and portions of the EGS, have been targeted by deep HST/WFC3 observations out to 1.6$\mu$m. These observations, which, in some cases, date back to 2010, can be used to explore potential proper motion in these sources, which would rule out them being extragalactic in origin. We only detect 8 sources in the WFC3 data, five in GOODS-S, two in GOODS-N, and one in CEERS, and we observe proper motion in only six. We do not observe a proper motion for two sources, JADES-GS-BD-6 or JADES-GN-BD-2. 

Using the {\tt photutils} centroids package, we calculated the center of the source in both the HST and the JWST observations independently, and we plot the thumbnails in Figure \ref{fig:BD_Proper_Motions} with circles indicating the recovered centroid positions. In addition, we provide the measured difference in position, which ranges from 0.12$^{\prime\prime}$ for JADES-GS-BD-1 to 0.75$^{\prime\prime}$ for JADES-GN-BD-3. We queried the MAST HST archive through \texttt{astroquery} package and obtained the HST WFC3 observation time of these brown dwarf candidates. We found that the mean time difference between the HST and JWST observations were 9--11 years. If we use those mean observed times for the individual sources, we derive proper motions of $0.013^{\prime\prime}$/yr for JADES-GS-BD-1, $0.013^{\prime\prime}$/yr for JADES-GS-BD-7, $0.017^{\prime\prime}$/yr for JADES-GS-BD-10, $0.050^{\prime\prime}$/yr for JADES-GS-BD-11, $0.074^{\prime\prime}$/yr for JADES-GN-BD-3, and $0.022^{\prime\prime}$/yr for CEERS-EGS-BD-3. 

We also searched for proper motions by comparing the JADES observations made in late 2022 to more recent medium-band NIRCam observations of the JADES Origins Field \citep[JOF, PID 3215; ][]{eisenstein2023b}. These data were reduced in a similar manner to that of the JADES data as described in Section \ref{sec:jades}, including the derivation of an astrometric solution based on Gaia stars. Because of the size of the JOF, we only detected two of our candidates in these observations, JADES-GS-BD-2 and JADES-GS-BD-5. For JADES-GS-BD-2, we don't detect an offset between the NIRCam F300M position measured from the JOF data and the F277W position measured from the JADES observations. For JADES-GS-BD-5, there is an 0.05$^{\prime\prime}$ difference in the NIRCam F300M position measured from the JOF data and the F277W position measured from the JADES observations, implying a $0.05^{\prime\prime}$/yr proper motion for the source.

We additionally searched through the Spitzer Space Telescope archive to see if any of our candidate brown dwarfs were detected in observations using IRAC instrument taken as part of the GOODS Spitzer Legacy program \citep{dickinson2003}. We specifically looked at the images taken in the IRAC2 (4.5$\mu$m) channel, as, from the SEDs, our candidates are brightest at this wavelength and therefore more likely to be detected. Only two sources were detected, JADES-GN-BD-3 and JADES-GS-BD-5. As JADES-GN-BD-3 is also observed in the HST WFC3 mosaic, we will use this measured separation for the proper motion measurement owing to the resolution of HST. We should note that the westward direction of the observed motion for JADES-GN-BD-3 is the same between the HST and Spitzer observations. 

For JADES-GS-BD-5, this source is too faint at $<2\mu m$ to be observed in the WFC3 data. To measure the offset in the Spitzer IRAC2 observation, we used the same method as was used to measure the HST WFC3 offsets, but we had to correct for the fact that the IRAC2 world coordinate solution (WCS) is offset from the Gaia-derived frame used in the JADES mosaics. To estimate this correction, we found the offset for the centroid of a nearby bright star, and subtracted this offset from the measured Spitzer centroid position. In Figure \ref{fig:BD_Proper_Motions}, we show JADES-GS-BD-5 with both the measured Spitzer position of the source as well as the corrected position, and derive a difference between the two positions of 0.62$^{\prime\prime}$. Given the positional accuracy of the Spitzer IRAC instrument ($<0.3^{\prime\prime}$ as per the Spitzer IRAC Knowledgebase\footnote{https://irsa.ipac.caltech.edu/docs/knowledgebase/spitzer\_irac.html}), this is a $>2\sigma$ offset. The date range provided for this Spitzer observation is August 12 2004 to August 18 2004, which would indicate a proper motion for JADES-GS-BD-5 of $\sim0.033^{\prime\prime}$/yr, which is slightly lower than the value we measured based on the NIRCam observations, but given the uncertainty in our WCS correction, these values are in agreement. More importantly, the direction of the offset (south, and very slightly to the west), is almost identical between the two independent measurements of the proper motion. 

\subsection{Discussion of Individual Sources} 

Because of the number of candidate sources in our sample, below, we will discuss each source individually, highlighting any previous discussion of these objects in the literature. For each source, we present an approximate subdwarf classification based on the best-fit effective temperatures following \citet{burgasser2002}. The true classification requires spectroscopic follow-up of these sources. 

\begin{deluxetable*}{l|ccccc|cccc|cccc}[ht!]
\tabletypesize{\footnotesize}
\tablecolumns{17}
\tablewidth{0pt}
\tablecaption{Derived Properties of the Brown Dwarf Candidates \label{tab:bd_derived_properties}}
\tablehead{
 \multicolumn{1}{c}{} & \multicolumn{5}{c}{Sonora Cholla} & \multicolumn{4}{c}{Sonora Bobcat} & \multicolumn{4}{c}{ATMO 2020}\\
 \colhead{Object ID} & \colhead{T$_{\mathrm{eff}}$ (K)} & \colhead{log(g)} & \colhead{log(Kzz)} & \colhead{R (R$_{\odot}$)} & \colhead{D (pc)}
 & \colhead{T$_{\mathrm{eff}}$ (K)} & \colhead{log(g)} & \colhead{R (R$_{\odot}$)} & \colhead{D (pc)}
 & \colhead{T$_{\mathrm{eff}}$ (K)} & \colhead{log(g)} & \colhead{R (R$_{\odot}$)} & \colhead{D (pc)}}
 \startdata
JADES-GS-BD-1 & 1150 & 5.5 & 2 & 0.077 & 749 & 1100 & 5.25 & 0.086 & 820 & 1100 & 5.5 & 0.125 & 1162 \\
JADES-GS-BD-2 & 1200 & 5.5 & 4 & 0.078 & 1223 & 1200 & 5.25 & 0.087 & 1460 & 1200 & 5.5 & 0.079 & 1295 \\
JADES-GS-BD-3 & 900 & 5.25 & 4 & 0.085 & 2274 & 1000 & 5.0 & 0.094 & 3480 & 900 & 5.5 & 0.078 & 2351 \\
JADES-GS-BD-4 & 500 & 3.505 & 4 & 0.128 & 1181 & 525 & 5.25 & 0.081 & 1230 & 500 & 5.5 & 0.101 & 1364 \\
JADES-GS-BD-5 & 500 & 5.5 & 2 & 0.073 & 251 & 375 & 4.0 & 0.114 & 190 & 350 & 5.0 & 0.104 & 160 \\
JADES-GS-BD-6 & 1000 & 5.5 & 4 & 0.076 & 1633 & 1000 & 5.0 & 0.094 & 2290 & 900 & 5.5 & 0.078 & 1549 \\
JADES-GS-BD-7 & 1000 & 5.5 & 2 & 0.076 & 770 & 1000 & 5.5 & 0.077 & 880 & 1000 & 5.5 & 0.13 & 1467 \\
JADES-GS-BD-8 & 500 & 3.505 & 4 & 0.128 & 804 & 500 & 5.0 & 0.089 & 840 & 500 & 5.5 & 0.101 & 949 \\
JADES-GS-BD-9 & 900 & 5.5 & 4 & 0.076 & 1429 & 900 & 5.0 & 0.093 & 2080 & 800 & 5.5 & 0.097 & 1710 \\
JADES-GS-BD-10 & 1050 & 3.748 & 4 & 0.141 & 380 & 1100 & 5.25 & 0.086 & 260 & 1100 & 5.5 & 0.125 & 377 \\
JADES-GS-BD-11 & 750 & 5.5 & 7 & 0.075 & 498 & 750 & 5.25 & 0.083 & 770 & 700 & 5.5 & 0.085 & 668 \\
JADES-GN-BD-1 & 650 & 3.748 & 4 & 0.129 & 688 & 700 & 5.0 & 0.091 & 740 & 700 & 5.5 & 0.085 & 696 \\
JADES-GN-BD-2 & 800 & 5.0 & 4 & 0.093 & 1137 & 850 & 5.5 & 0.076 & 1250 & 800 & 5.5 & 0.097 & 1407 \\
JADES-GN-BD-3 & 750 & 5.5 & 2 & 0.075 & 335 & 600 & 5.25 & 0.082 & 290 & 600 & 5.5 & 0.1 & 354 \\
CEERS-EGS-BD-1 & 800 & 5.25 & 4 & 0.084 & 1682 & 850 & 5.25 & 0.084 & 2330 & 800 & 5.5 & 0.097 & 2353 \\
CEERS-EGS-BD-2 & 700 & 5.25 & 4 & 0.083 & 965 & 650 & 5.5 & 0.074 & 1020 & 600 & 5.5 & 0.1 & 1149 \\
CEERS-EGS-BD-3 & 1100 & 5.5 & 4 & 0.077 & 642 & 1100 & 5.0 & 0.095 & 870 & 1000 & 5.5 & 0.13 & 984 \\
CEERS-EGS-BD-4 & 1050 & 5.0 & 2 & 0.095 & 1391 & 1000 & 5.0 & 0.095 & 1350 & 1000 & 5.0 & 0.13 & 1813 \\
CEERS-EGS-BD-5 & 1000 & 3.505 & 4 & 0.145 & 4222 & 1100 & 5.25 & 0.086 & 3100 & 1100 & 5.5 & 0.125 & 4503 \\
CEERS-EGS-BD-6 & 600 & 5.0 & 2 & 0.09 & 1059 & 525 & 5.0 & 0.09 & 1130 & 500 & 5.0 & 0.101 & 1098 \\
CEERS-EGS-BD-7 & 650 & 3.505 & 2 & 0.133 & 1290 & 650 & 5.0 & 0.09 & 1080 & 600 & 4.5 & 0.1 & 1006 \\
\enddata
\end{deluxetable*}

\subsubsection{JADES-GS-BD-1} 
This source is very bright, with fluxes above 100 nJy in five of the JADES filters, and the Sonora Cholla fit accurately matches the observed fluxes. The fit indicates that the source is at T$_{\mathrm{eff}} = 1150$K (T3-T5), and a distance of $\sim 750$ pc. This source is unresolved, and has an observed separation between the HST WFC3 F160W image and the JWST NIRCam F150W image of $0.12^{\prime\prime}$. Given the HST detection, it is puzzling that this source does not appear to have been written about: a search through the literature for any sources within $5^{\prime\prime}$ of the WFC3 position of this object did not result in any matches. 

\subsubsection{JADES-GS-BD-2} 
This source is a very bright candidate (F410M flux of $\sim 180$ nJy) detected significantly at all of the JADES NIRCam wavelengths, with obvious diffraction features in the 3 - 4$\mu$m images, suggesting that it is unresolved. The larger $\chi^2$ of the Sonora Cholla fit is due in large part to the mismatch in the F150W filter, while the resulting fit does accurately match the observed fluxes, indicating that this is potentially an early T dwarf at 1.2 kpc. The {\tt EAZY} galaxy template fit does not accurately predict the relatively fainter F277W and F335M fluxes in this source, attributing the LW emission to 3.3$\mu$m PAH emission. 

\subsubsection{JADES-GS-BD-3}  
This brown dwarf candidate has an excellent Sonora Cholla fit with T$_{\mathrm{eff}} = 900$K (T5-T6) and a derived distance of 2.2 kpc. The Sonora Bobcat fit has a slightly higher temperature (1000 K), and results in a distance of almost 3.5 kpc, which would place it firmly in the Milky Way halo. The extremely blue F115W - F200W color and the lack of an F277W detection make it difficult to conclude that the source is extragalactic.  

\subsubsection{JADES-GS-BD-4} 
This brown dwarf candidate is very faint at wavelengths shorter than 3.5$\mu$m, with fluxes of only $\sim$2 nJy in F115W and F150W, and no significant detections in F200W or F277W. The Sonora Cholla fit for this source indicates that it is only a 500 K (Y dwarf) source at a distance of almost 1.2 kpc, and this is one of three sources where the Sonora Cholla fit results in such a low temperature. The {\tt EAZY} redshift ($z_a = 11.85$) is ruled out due to the detection of F115W flux shortward of the best-fit Lyman break. In the absence of detections at short wavelengths, this source would appear to be a dusty red galaxy or even an F277W dropout at $z > 20$. Faint Y-dwarfs should be explored as possible alternate explanations for ultra high-redshift JWST galaxy candidates. 

\subsubsection{JADES-GS-BD-5} 
This brown dwarf candidate has a Sonora Cholla fit with $\chi^2 = 1320$, but this is driven by the small uncertainties on the fluxes, the blue F277W - F335M color in this source (it is the only object in our sample not selected by the F115W-F277W-F444W criteria shown in the middle panel of Figure \ref{fig:BD_Color_Color_All}), and the relative lack of flux at $< 1.6\mu$m. The best-fit results in a low effective temperature (500 K, a Y dwarf) and positions the source at only 251 pc away, the closest among the entire sample. As discussed earlier in this section, this source has an observed proper motion between the Spitzer IRAC2 and JWST NIRCam F444W observations of $0.62^{\prime\prime}$ after correcting for the different reference frames. This best galaxy template {\tt EAZY} fit returns a photometric redshift of $z_a = 17.84$, which is at odds with the observed F115W, F150W, and F200W fluxes, and the $\chi^2$ is still significantly worse than the Sonora Cholla fit. This could result from a limitation of the cloud-free Sonora Cholla models, as any clouds, or differences in the molecular absorption from the models, would significantly change the shape of the SED. While the source could be a binary brown dwarf system \citep[see][for examples in the literature]{burgasser2003b, cruz2004, burgasser2006, dupuy2015, kiwy2022, calissendorff2023}, this may not significantly effect the observed colors. 

\subsubsection{JADES-GS-BD-6} 
This source is quite well fit by a 1000 K Sonora Cholla model, indicating that it is potentially a T5-T6 dwarf at a distance of 1.6 kpc. The {\tt EAZY} galaxy template fits attempts to replicate the very blue F115W - F277W color with a Lyman break and the red F356W - F444W color with [OIII]$\lambda$5007 emission boosting the F410M and F444W fluxes, but the resulting fit is very poor. 

\subsubsection{JADES-GS-BD-7} 
The Sonora Cholla fit for this candidate indicates that it is at 1000K (T5 - T6), at a distance of $\sim$800 pc. The fit is significantly better than the lowest $\chi^2$ {\tt EAZY} fit, and the evidence for this source being within the Milky Way is strengthened by the $0.15^{\prime\prime}$ separation between the HST WFC3 F125W centroid and the JWST NIRCam F115W centroid. This source was discussed previously in \citet{oesch2012}, \citet{lorenzoni2013}, and \citet{finkelstein2015}.  \citet{oesch2012} removed the source from their list of $z \sim 7$ candidates as it had colors consistent with being a T-dwarf. It should be noted that the proper motion between the HST and JWST observations for this source was first discussed in \citet{bunker2023} under the ID 10035328. 

\subsubsection{JADES-GS-BD-8} 
This source is at SNR $> 10$ in the F115W, F356W, F410M and F444W filters, but has SNR $< 3$ in the other JADES NIRCam filters. The best Sonora Cholla fit indicates the source could possibly be a 500K Y-dwarf at 804 pc, while the best {\tt EAZY} fit indicates the source could be at $z = 12.46$, a redshift ruled out by the F115W detection. The coverage at 1.15 $\mu$m is very important for helping to confirm these faint Y-dwarfs. 

\subsubsection{JADES-GS-BD-9} 
This source is bright, with a point-like morphology and an excellent Sonora Cholla fit with T$_{\mathrm{eff}} = 900$K (T6 - T7) indicating a distance of 1.4 kpc. The poor {\tt EAZY} fit significantly overpredicts the 1.5 - 3.5$\mu$m flux for this source. 

\subsubsection{JADES-GS-BD-10} 
This candidate is the brightest source in our sample, with an F410M flux of $> 4000$ nJy. The Sonora Cholla fit agrees with the observed photometry quite well, and indicates that the source is potentially at 1050 K (T0 - T3), at a distance of 380 pc. The {\tt EAZY} fit is much worse, and while a $z_a = 6.89$ galaxy solution is possible, it overpredicts the F277W and F335M fluxes. This source has a proper motion of $0.017^{\prime\prime}$/yr, and so an extragalactic origin is unlikely. It has been very well studied in the literature owing to its brightness, and was first discovered in a search for $z \sim 6$ galaxies in GOODS-S in \citet{stanway2003}, where the authors suggest it is likely a star owing to its colors. Later, in \citet{vanzella2009}, the authors remove the source from a list of high-redshift candidate galaxies due to detection of proper motion.

\subsubsection{JADES-GS-BD-11} 
This brown dwarf candidate is near a gap in the short-wavelength JADES mosaic, as seen in the thumbnails in Figure \ref{fig:BD_SEDs_2}. The {\tt forcepho} flux measurements were not affected by the position of the source, and this source is quite bright ($\sim 200$ nJy at 4$\mu$m). The best Sonora Cholla fit returns T$_{\mathrm{eff}} = 750$K (T9) at $\sim$500 pc. This source has an observed offset of $0.56^{\prime\prime}$ between the HST WFC3 F125W and JWST NIRCam F150W positions. In addition, this source is GSDZ-2480745501 in \citet{bouwens2015}, with an $z_{\tt EAZY} = 6.69$, which would put the Lyman break at 0.93$\mu$m. The very red LW color (F335M - F444W = 4.07), and the observed proper motion call this interpretation into doubt. 

\subsubsection{JADES-GN-BD-1} 
This candidate is very faint at short wavelengths, and both bright and unresolved at longer wavelengths, with fluxes of around almost 200 nJy at F410M and F444W. The best fit from the Sonora Cholla models suggests that the candidate is only at 650 K (an early Y dwarf) at only $\sim700$ pc away. The {\tt EAZY} galaxy template fit is very poor, with a photometric redshift of 11.88 in disagreement with the significant F115W detection. 

\subsubsection{JADES-GN-BD-2} 
The Sonora Cholla best fit for this source is an 800K (T7 - T8 dwarf) model at 1 kpc, the furthest observed in GOODS-N. The {\tt EAZY} fit, which cannot account for the observed faint F200W, F277W, or F335M emission, attempts to fit the source as an F090W dropout with strong line emission, and has a significantly worse $\chi^2$. 

\subsubsection{JADES-GN-BD-3} 
This brown dwarf candidate in GOODS-N is detected in each of the JADES NIRCam bands, with $\sim700$ nJy fluxes in F410M and F444W. As a result, the uncertainties on the fluxes are quite small, leading to a large $\chi^2$ for the Sonora Cholla fit. However, from Figure \ref{fig:BD_SEDs_1}, it can be seen that the fit accurately matches the fluxes and the resulting parameters suggest that the source could be nearby (334 pc) with T$_{\mathrm{eff}} = 750$ K (Y dwarf). This source has a proper motion of $0.074^{\prime\prime}$/yr, the largest we observe in our sample. This lends strong credibility to this source being within our own Galaxy. 

A source at the HST location of this object was discussed in \citet{bouwens2015} and \citet{finkelstein2015} as a $z\sim7$ galaxy, likely due to the HST WFC3 observations only probing out to $1.6\mu$m. An analysis by \citet{endsley2021} of the Spitzer observations for this source noted its red 3.6 - 4.5 $\mu$m color, but it was not included in their final sample of $z \sim 7$ [OIII]+H$\beta$ emitters. 

\subsubsection{CEERS-EGS-BD-1} 
The Sonora Cholla model for this source has a best-fit T$_{\mathrm{eff}} = 800$ K (T7 - T8) at a distance of 1.7 kpc. The fit is significantly better than the {\tt EAZY} fit, which overpredicts both the F200W and F277W non-detections and the F814W flux. 

\subsubsection{CEERS-EGS-BD-2} 
The brown dwarf candidate has an excellent fit ($\chi^2 = 4$) with T$_{\mathrm{eff}} = 700$\,K (T8 - L0) and a distance of 970 pc. 

\subsubsection{CEERS-EGS-BD-3} 
This candidate is the brightest among the CEERS sources, with F410M flux of $\sim 480$ nJy. It is located north of a galaxy (separated by 1.4$^{\prime\prime}$) at $z_{\mathrm{spec}} = 0.82$ \citep{momcheva2016}. The Sonora Cholla fit indicates that the object could be at T$_{\mathrm{eff}} = 1100$\,K (T2 - T5) and a distance of 642 pc. We also observe a proper motion of 0.022$^{\prime\prime}$/yr for the source, demonstrating that this source is likely not extragalactic in nature. A search through the literature for sources detected at this HST-derived position did not return any matches, potentially due to the proximity to the nearby galaxy. 

\subsubsection{CEERS-EGS-BD-4} 
This brown dwarf candidate is the brightest across the CEERS footprint, with an F444W flux of 116 nJy, and the Sonora Cholla fit is significantly better than the {\tt EAZY} fit ($\chi^2 = 237$). The best fit indicates a T$_{\mathrm{eff}} = 1050$ K (T3 - T5), with a distance of 1.4 kpc. In this source, and multiple other CEERS sources, the {\tt EAZY} fit predicts a flux at $< 1 \mu$m that is not observed in the HST ACS F814W mosaics. 

\subsubsection{CEERS-EGS-BD-5} 
This Sonora Cholla fit to this source has a $\chi^2 = 2.7$, T$_{\mathrm{eff}} = 1000$ K (T4 - T6), and most importantly, a predicted distance of 4.2 kpc, making it the farthest brown dwarf candidate in our sample. The {\tt EAZY} fit to the NIRCam data predicts an F814W detection that we do not observe (the $0.2^{\prime\prime}$ diameter circular aperture flux for this source is $-0.3 \pm 1.8$ nJy, consistent with zero). If this source were a galaxy, it would need to have a strong hot dust component coupled with a very blue UV slope ($\beta = -3.75^{0.25}_{0.28}$). At this slope, it would be a significant outlier among the JADES sample of F090W dropout galaxies assembled by \citet{topping2023}.

\subsubsection{CEERS-EGS-BD-6}
This source is not significantly detected in F200W or F277W, and is only detected at SNR $= 5.1$ in F115W and SNR $= 2.9$ in F150W. The best-fit indicate a T$_{\mathrm{eff}} = 600$ K (Y dwarf) at 1.1 kpc. The {\tt EAZY} fit, at $z_a = 12.0$, is not in agreement with the observed F115W flux. 

\subsubsection{CEERS-EGS-BD-7} 
The best-fit parameters for the Sonora Cholla fit to the photometry for this source indicate a T$_{\mathrm{eff}} = 650$K (late T or early Y dwarf) at 1.3 kpc. The fit, however, underpredicts the observed F150W, F200W, and F277W fluxes. 

\section{Discussion} \label{sec:discussion}

In this section, we will discuss the implications of these brown dwarf candidates by exploring their positions within the Milky Way and distances from the Sun, as well as their observed colors with an eye towards how these types of objects can be detected in current and future deep extragalactic catalogs. 

\subsection{Position in the Milky Way} \label{sec:milkywayposition}

\begin{figure}
  \centering
  \includegraphics[width=0.5\textwidth]{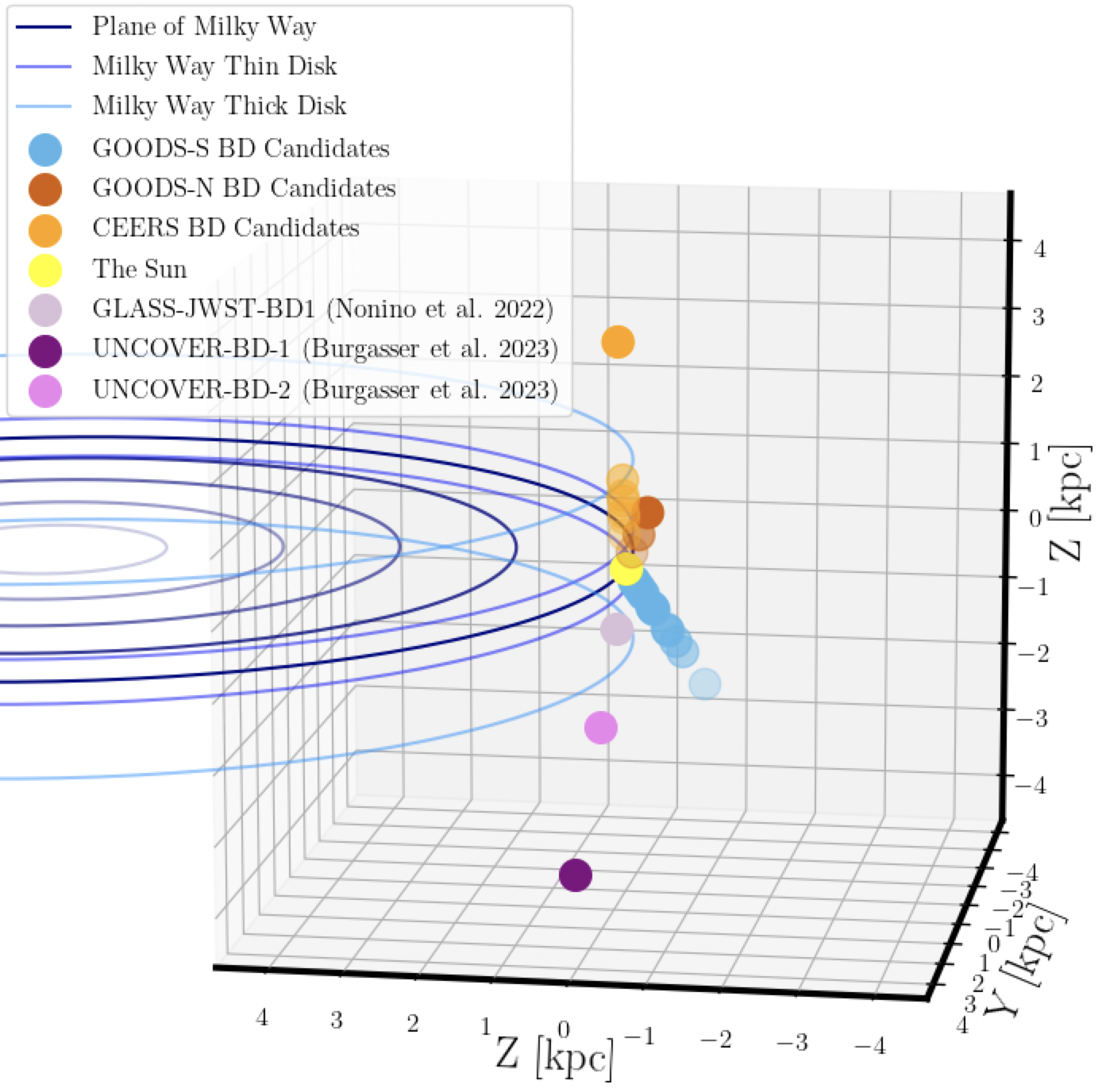}
  \caption{Positions of the JADES and CEERS brown dwarf candidates, with JADES GOODS-S sources in shades of blue, JADES GOODS-N sources in shades of red, and CEERS sources in shades of orange, plotted in three dimensional space with respect to the Milky Way Galaxy disk. The darker the shade, the closer the source is to the viewpoint of the plot. The distances for these sources are from fits using the Sonora Cholla models along with the solar metallicity Sonora Bobcat models. We show the central plane of the Milky Way with a navy blue line, and then the thin and thick disk with lighter blue lines. We also plot the Sun as a yellow circle, and the spectroscopically-confirmed brown dwarfs from the UNCOVER and GLASS surveys \citep{nonino2023,langeroodi2023,burgasser2023} with purple points. For these sources, we adopt the distances derived in \citet{burgasser2023}. 
  \label{fig:milky_way_position}}
\end{figure}

In Figure \ref{fig:milky_way_position}, we plot the positions of our brown dwarf candidates with respect to the Sun and our Galaxy. For this figure, we adopt the distances measured using the Sonora Cholla models using the solar metallicity Sonora Bobcat evolutionary models given in Table \ref{tab:bd_derived_properties}. We plot the GOODS-S sources with blue points, the GOODS-N sources with red points, and the CEERS sources with orange points. We include the \citet{nonino2023} GLASS brown dwarf, as well as the two UNCOVER brown dwarfs discussed in \citet{langeroodi2023} and \citet{burgasser2023}, with purple points. For these three sources, their distances are taken from \citet{burgasser2023}. We also plot the plane of the Milky Way with blue curves. The darkest blue curve represents the central plane of the Galaxy, with the thin disk (300 pc scale height) and thick disk (1.4 kpc scale height) represented with lighter blue curves. The bulk of the JADES candidates lie within the thick disk, although there are four that may be outside the thick disk in the Galactic halo (JADES-GS-BD-6, JADES-GS-BD-3, CEERS-EGS-BD-5, and CEERS-EGS-BD-1). It should be cautioned that the distance estimates from the fitting are highly uncertain. 

\begin{figure}
  \centering
  \includegraphics[width=0.48\textwidth]{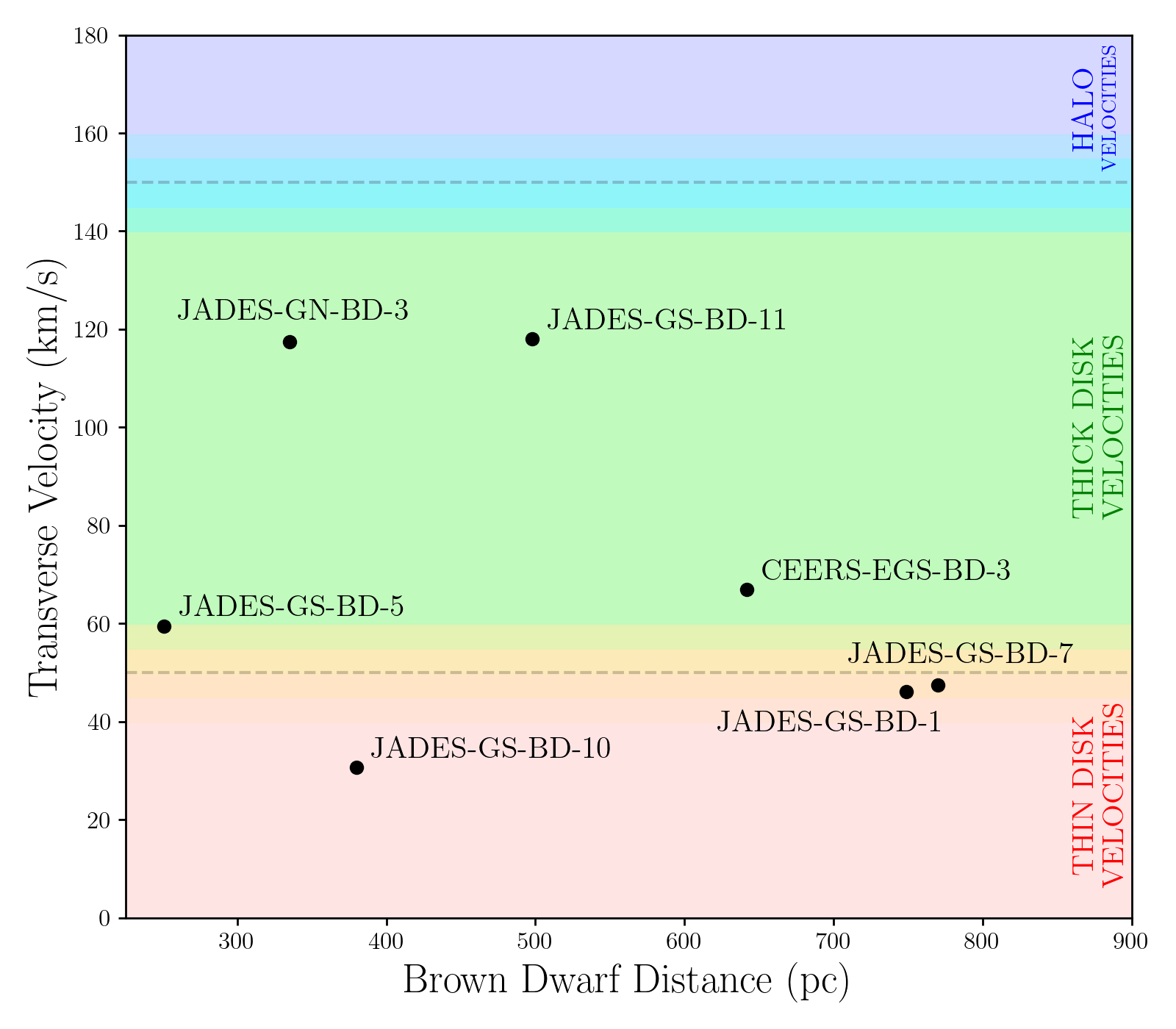}
  \caption{Transverse velocities of the brown dwarf candidates plotted against their estimated distances. We also show, with colored bands and dashed lines, the approximate transverse velocity ranges of opbserved thin-disk objects (0 - 50 km/s), thick-disk objects (50 - 150 km/s), and halo objects ($> 150$ km/s). The bulk of the brown dwarf candidates with proper motions in our sample have transverse velocities similar to thick-disk objects. 
  \label{fig:transverse_velocities}}
\end{figure}

To better explore the nature of these brown dwarf candidates, we calculated their transverse velocities from their proper motions and distances using the equation $v_T = 4.74\mu d$ (here, $v_T$ is the transverse velocity in km/s, $\mu$ is the proper motion in arcseconds per year, and $d$ is the distance in parsecs). We estimate transverse velocities with respect to the Sun of 46 km/s for JADES-GS-BD-1, 59 km/s for JADES-GS-BD-5, 47 km/s for JADES-GS-BD-7, 31 km/s for JADES-GS-BD-10, 118 km/s for JADES-GS-BD-11, 118 km/s for JADES-GN-BD-3, and 67 km/s for CEERS-EGS-BD-3. These values are, on average, higher than more local brown dwarfs at similar temperatures from \citet{kirkpatrick2021}, where most have transverse velocities below 50 km/s. We plot these values against our estimated distances in Figure \ref{fig:transverse_velocities}. On this Figure, we indicate  ranges in transverse velocity corresponding to the thin disk (0 km/s - 50 km/s), thick disk (50 km/s - 150 km/s) and halo ($> 150$ km/s) populations with colored bands \citep{casertano1990, robin2003, nissen2004}. While the transverse velocities for the sources in our sample with proper motion agree with those in the thin and thick disks, their distances in combination with these velocities provide strong evidence that these sources are members of the thick disk of the Galaxy. 

Several authors have explored the number density of low-mass stars and brown dwarfs in the Milky Way from both observations and simulations. \citet{ryanreid2016} provided predicted number densities for ultracool dwarfs between spectral types M8-T8 for thin and thick disk exponential models motivated by estimates of their luminosity function. These authors predict a number density of T0-T5 dwarfs at all observed magnitudes (combining the thick and thin disk estimates) in GOODS-S of $3.43\times10^{-2}$ arcmin$^{-1}$, in GOODS-N of $3.32\times10^{-2}$ arcmin$^{-1}$, and in the EGS of $3.84\times10^{-2}$ arcmin$^{-1}$. Given the current JADES area of 67 arcmin$^2$ for GOODS-S, 58 arcmin$^2$ for GOODS-N, and 90.8 arcmin$^2$ in the CEERS area in the EGS, the \citet{ryanreid2016} prediction would be 2.3 T-dwarfs across the JADES GOODS-S footprint, 1.9 across GOODS-N, and 3.5 across CEERS. If we define the T0 - T5 dwarf temperature range as T$_{\mathrm{eff}} \sim 1000 - 1300$K \citep{burgasser2002}, then we find only five candidates in GOODS-S with best fits in this temperature range, none in GOODS-N, and 3 in the CEERS EGS footprint. This is in statistical agreement with the combined prediction across GOODS-S, GOODS-N, and CEERS EGS from \citet{ryanreid2016}. 

An alternate method of comparing the observed number density of brown dwarf candidates is adopted by \citet{wang2023}, where these authors compare the likelihood of finding their source, CEERS-BD1, with the the observed space density of early-L through early-Y dwarfs from \citet{kirkpatrick2021}. Table 15 in \citet{kirkpatrick2021} provides space densities as a function of T$_{\mathrm{eff}}$ complete only out to 20 pc. If we co-add the number densities of T0 to Y1.5 sources from this table, these authors estimate a total of $14.92\times10^{-3}$ arcmin$^{-1}$. which, If we multiply this estimate by the total area of the JADES GOODS-S and GOODS-N and CEERS EGS footprints (216 arcmin$^2$), implies only 3.2 brown dwarfs across this entire temperature range. The bulk of the candidates in our sample have best-fit distances outside the thin disk probed by the \citet{kirkpatrick2021} sample, however. Indeed, these results and others in the literature can be added to the growing number of substellar candidates found in extragalactic deep fields as discussed in the Introduction. 

\subsection{Comparing to High-Redshift Galaxy Colors} \label{sec:highzgalaxycolors}

\begin{figure}
  \centering
  \includegraphics[width=0.5\textwidth]{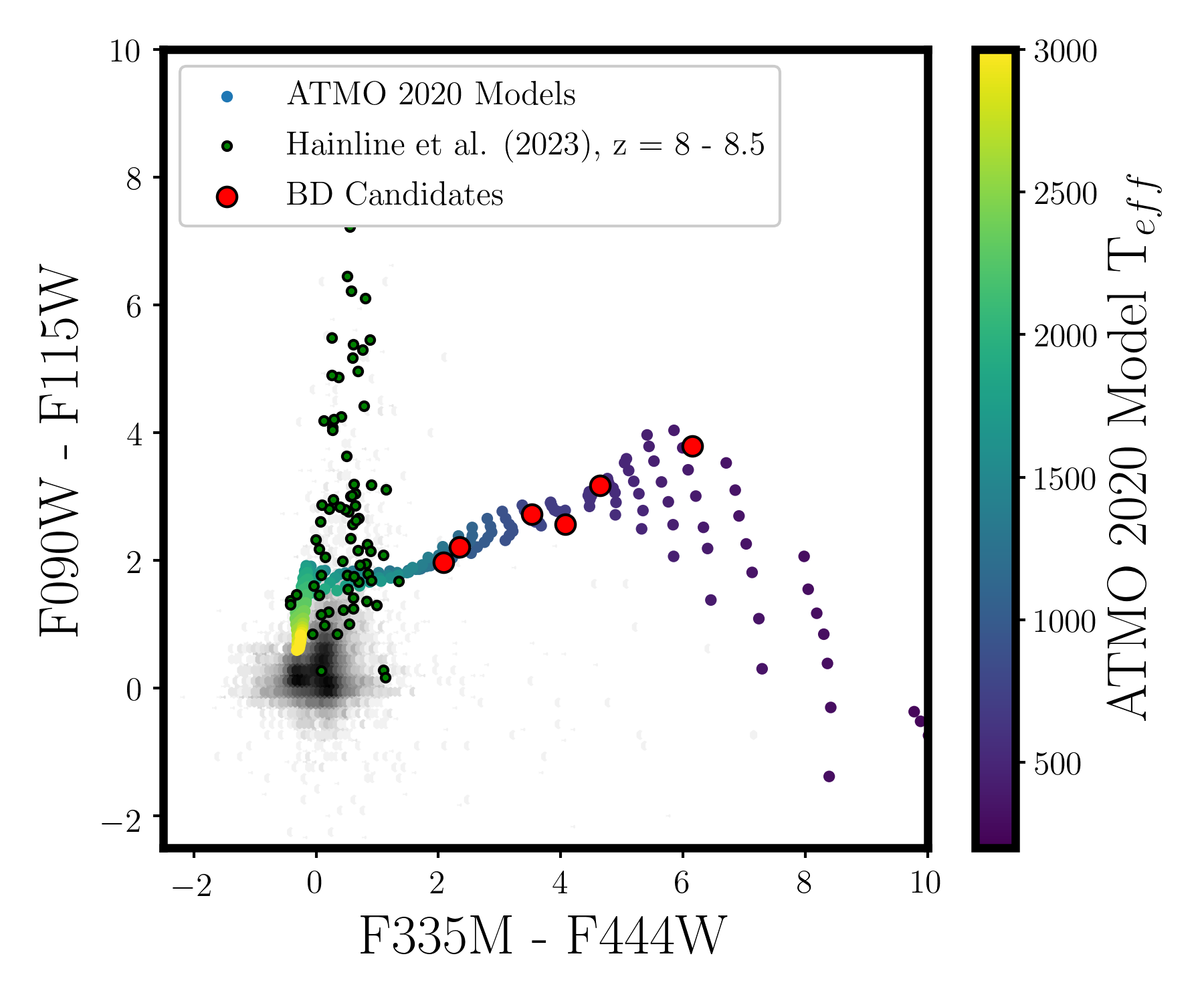}
  \caption{F090W-F115W vs F335M-F444W color. In this plot, we show those JADES sources with m$_{F444W} < 28.5$ in grey, but we plot the brown dwarf candidates detected in F090W, F115W, F335M, and F444W with red circles, as well as the ATMO 2020 model colors with points signifying their effective temperature as shown in the colorbar. In addition, we also plot a sample of F090W Lyman dropout galaxies from \citet{hainline2023} with {\tt EAZY} photometric redshifts between $z = 8 - 8.5$. A red F335M - F444W color is helpful for distinguishing between F090W dropout galaxies at $z \sim 8$ and low-temperature brown dwarfs, although at higher temperatures, the ATMO models predict that late M and early L dwarfs would have similar colors to the dropout galaxies. 
  \label{fig:BD_Color_vs_highz}}
\end{figure}

As discussed in Section \ref{sec:results}, some of the brown dwarf candidates in our sample have previously appeared in the literature as potential high-redshift galaxies owing to them being selected as dropout galaxies at 0.9$\mu$m and the limited HST WFC3 wavelength coverage in the near-IR. Following work exploring the use of the 3 - 4$\mu$m slope in separating brown dwarf candidates from F090W dropous in \citet{hainline2020}, in Figure \ref{fig:BD_Color_vs_highz} we plot the F090W - F115W colors vs. the F335M - F444W colors for the JADES survey objects with m$_{F444W} < 28.5$. Because the CEERS survey does not include coverage in F090W or F335M, we do not plot any CEERS sources. We also plot the six brown dwarf candidates in our sample with detections in F090W, F115W, F335M, and F444W. As a subsample of the high-temperature Sonora Cholla models do not extend to wavelengths that cover the entirety of the F090W filter, we instead include the predictions from ATMO 2020, with points colored by the effective temperature of the model. The F335M fluxes are quite low in our candidates, which, if these were brown dwarfs, would be caused by molecular absorption by CH$_{4}$ at 3 - 3.6$\mu$m.  

We additionally plot a sample of $z = 8.0 - 8.5$ galaxy candidates from \citet{hainline2023}. In this redshift range, these sources are are F090W Lyman dropouts. Similar to the high-z sample, we measure F090W-F115W $> 1$ for the brown dwarf candidates (five of the six have F090W-F115W $> 2$). The low-temperature brown dwarf candidates differentiate themselves from the high-$z$ sample by having F335M - F444W $> 2.0$. While ionized emission (e.g. [OIII]$\lambda$5007) can provide extra flux in the F410M and F444W filters, there are no galaxies in the $z = 8.0 - 8.5$ JADES sample with F335M - F444W colors similar to that of the low-mass brown dwarfs. At higher temperatures, the predicted ATMO 2020 colors for late M and early Y dwarfs are similar to the F090W dropouts, however. 

In \citet{labbe2023b}, the authors selected sources red LW colors from the JWST Ultradeep NIRSpec and NIRCam ObserVations before the Epoch of Reionization survey \citep[UNCOVER]{bezanson2022}. A number of the members of their sample have colors similar to the brown dwarf candidates selected here. The authors fit these sources with both galaxy stellar light as well as reprocessed light from dust heated by accretion onto a supermassive black hole. While these fits can account for the red F335M - F444W colors in these sources, they have difficulties fitting the extremely blue F115W - F277W colors at the same time. The alternate explanation, that a subsample of their sources are unresolved brown dwarf candidates, was recently confirmed by independent reductions of spectroscopy for two of their sources in \citet{langeroodi2023} and \citet{burgasser2023}. Multiple authors have looked at similar samples of extremely red sources in extragalactic datasets \citep[e.g. ][among others]{barro2023, perezgonzalez2023, franco2023}, and we have outlined color selection criteria here which may be used to find the brown dwarfs from among those samples.

If we assume that the candidates in our sample are F090W dropouts at $z = 8.0 - 8.5$, we can measure a potential UV slope for the sources by performing a linear fitting to the F115W, F150W, and F200W fluxes. The resulting slopes, which range between $\beta = -3.62$ (JADES-GS-BD-5) and $\beta = -6.75$ (JADES-GS-BD-8), have a median $\beta = -4.81$. This is significantly bluer than the assembled F090W dropouts selected across the JADES survey in \citet{topping2023}, which have a median $\beta = -2.32^{+0.03}_{-0.02}$. This provides more evidence that these sources are not extragalactic in origin. 

\subsection{Brown Dwarf Selection in Other Extragalactic Programs} \label{sec:otherprograms}

The JADES depth and filter selection, along with the large number of candidate brown dwarfs in our sample, allows us to predict number counts of sources found in current and future extragalactic JWST programs. In particular, we can compare our sources to the survey depths for the COSMOS-Web \citep{casey2022} and PRIMER \citep{dunlopprimer2021} programs which will feature NIRCam photometric coverage across large areas. For COSMOS-Web, the total area will be 1929 square arcmin with 5$\sigma$ point-source depths of 26.9 - 28.3 across four NIRCam filters (F115W, F150W, F277W, and F444W). PRIMER will survey 378 square arcminutes with eight filters (F090W, F115W, F150W, F200W, F277W, F356W, F410M, and F444W) down to 5$\sigma$ point-source depths of 27.6 - 29.5. Because of these extended areas, these surveys should uncover large numbers of these relatively rare brown dwarf candidates similar to those discussed in this paper. 

\begin{figure*}
  \centering
  \includegraphics[width=0.98\textwidth]{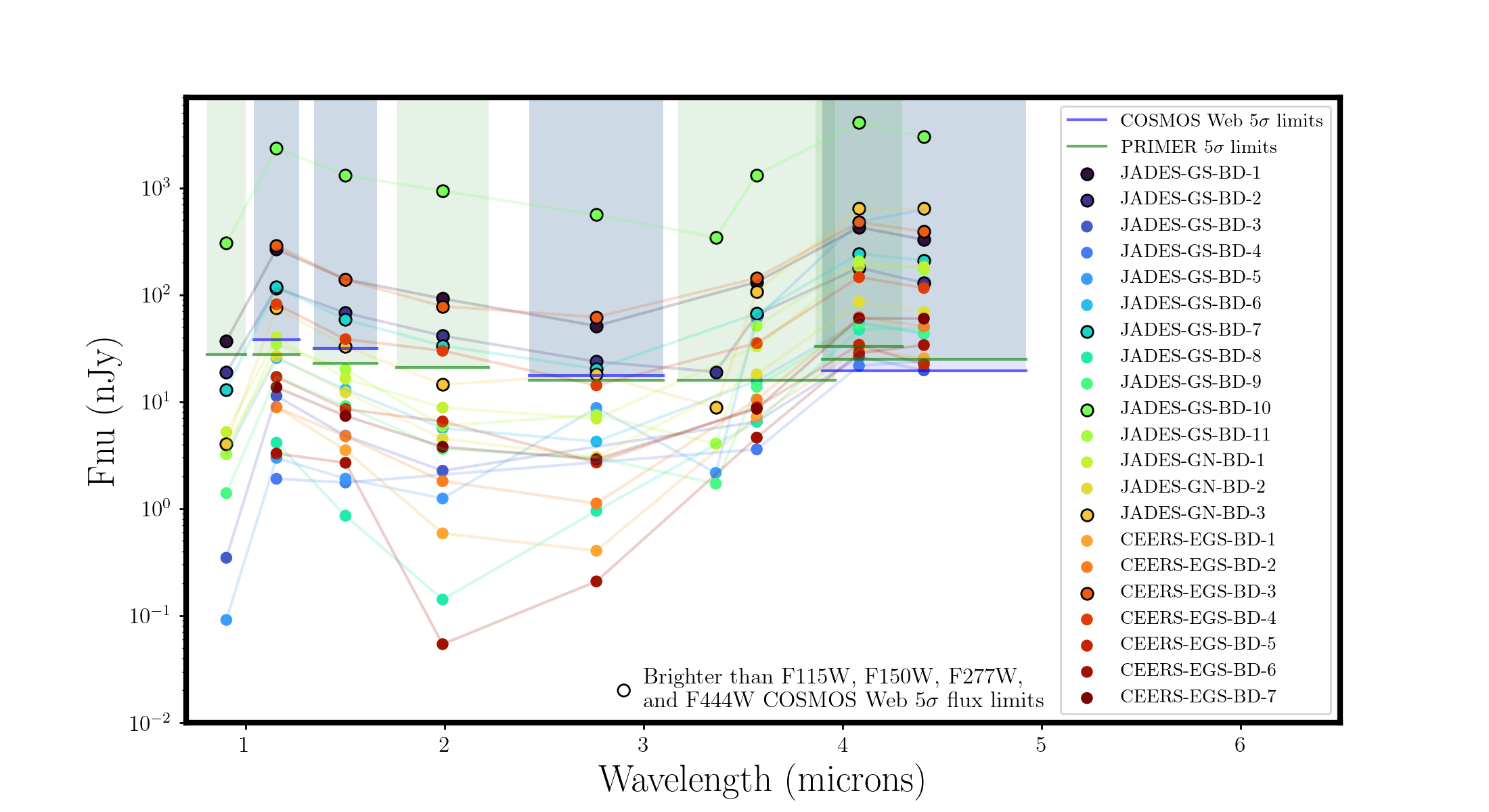}
  \caption{Brown dwarf candidate SEDs plotted along with the 5$\sigma$ point source depths for the COSMOS Web (blue shaded area) and Primer (green shaded area) surveys. We highlight those sources that would be detected above the COSMOS Web limits in F115W, F150W, F277W, and F444W with black heavy circles. While our sources are bright enough to be detected in F444W and F410M (for PRIMER), at $< 3\mu$m the majority are too faint, making characterizing their properties difficult in these large area programs.  
  \label{fig:BD_Survey_Depths}}
\end{figure*}

In Figure \ref{fig:BD_Survey_Depths}, we plot the SEDs for our sources and indicate the 5$\sigma$ point-source depths for both the PRIMER and COSMOS-Web filters to demonstrate which of the sources in our sample would be found in these surveys. For COSMOS-Web, we use the values estimated for the deepest portion (4 exposures) of the full survey as measured using $0.15^{\prime\prime}$ radius apertures and provided in \citet{casey2022}. For PRIMER, we use the measurements for the current data using $0.16^{\prime\prime}$ radius apertures and provided in \citet{morishita2023}, but we note that this is shallower than the future depths for the completed survey. 

Because of how our sources were chosen with $m_{\mathrm{F444W}} < 28.5$, each of the sources would be significantly detected in that filter for both PRIMER and COSMOS-Web. However, for the majority of our sources, they are not bright enough at short wavelengths to be observed in these shallow, large area surveys, so the actual selection of these sources would be difficult. When we compare to the COSMOS-Web depths and filters, we do find that six of the sources in our sample have F115W, F150W, and F277W fluxes brighter than the 5$\sigma$ flux limits, and five of these sources have best-fit T$_{\mathrm{eff}} > 1000$K. The remaining object, JADES-GN-BD-3, has T$_{\mathrm{eff}} = 750$K, demonstrating that bright sources with these low effective temperatures, while rare, are still accessible to this survey. We highlight these sources with thick outlined points in Figure \ref{fig:BD_Survey_Depths}. One additional source, CEERS-EGS-BD-4, is just fainter than the F277W flux limit for COMOS-Web. Across the JADES+CEERS footprints we find only $2.7\times10^{-2}$ sources arcmin$^{-2}$, which which would result in a total of 52 brown dwarfs bright enough to be detected in all four filters across the COSMOS-Web survey footprint, although this is a rough estimate given the limited number of sources we find. For the current depths of the PRIMER survey the number of sources that satisfy the detection threshold in F115W, F150W, and F277W from our sample is only four, although two additional sources would be detected in F115W and F150W, but not F277W. For the PRIMER depths, then, our sample predicts $1.8\times10^{-2}$ sources arcmin$^{-2}$, and a total of 6.8 brown dwarfs bright enough to be detected in F115W, F150W, F277W, and F444W, although this is at the current depths, which will increase with future observations. 

While all of our sources will still be found at 4$\mu$m at the COSMOS-Web and PRIMER survey depths, to characterize the SEDs of the cooler Y-dwarfs requires deeper observational data targeting primarily the 1 - 2$\mu$m wavelength range. With such large areas probed, the COSMOS-Web and PRIMER surveys still represent an incredible opportunity for understanding statistical samples of low-mass T-dwarfs and rare bright Y-dwarfs at hundreds of parsec or even kiloparsec distances. 

\section{Conclusions} \label{sec:conclusion}

We have searched the first year of deep NIRCam JADES and CEERS extragalactic photometric data to hunt for low-mass brown dwarf candidates. These sources are vital for understanding the low-mass end of the initial mass function, especially for the types of low-metallicity stars that are found in the Milky Way thick disk and halo. Our primary conclusions are:

\begin{enumerate}
    \item Using color selection and visual inspection, we found a sample of 21 sources across the combined 215.8 square arcminutes (0.09 candidates arcmin$^{-2}$). These sources were chosen by targeting the blue F115W - F277W and red F277W - F444W colors observed in late L, T, and Y subdwarfs.
    \item After fitting these sources to the \citet{karalidi2021} Sonora Cholla models, we find that these objects range in temperature between T$_{\mathrm{eff}} = 500 -1200$K, and are at distances of $0.1 - 4.2$ kpc. While the majority of the sources are in the thick disk, we do find four sources that are potentially in the Milky Way halo, based on their distances. These sources are poorly fit with galaxy template models. We find similar results when fitting with the \citet{phillips2020} ATMO 2020 models, albeit at larger distances. 
    \item By comparing these sources to existing HST/WFC3 and Spitzer infrared data, we estimate proper motions for seven of the candidates of $0.013^{\prime\prime}$/yr - $0.074^{\prime\prime}$/yr, providing evidence that these sources are not extragalactic in nature. The directions of these proper motions are roughly in line with the axis of the Milky Way. 
    \item We present a set of color criteria which can help separate these brown dwarf candidates from NIRCam F090W dropouts at $z \sim 8 - 8.5$, which do not have the very red F335M - F444W colors seen in subdwarfs.
    \item We use our sample to explore the temperatures of brown dwarfs that might be observed in the PRIMER and COSMOS Web surveys, which probe much larger areas than JADES and CEERS at shallower depths. Based on comparing to the fluxes of our brown dwarf candidates, we find that sources at T$_{\mathrm{eff}} < 1000$K are too faint at $<3\mu$m to be seen in these surveys, and will appear only in the F410M and F444W filters. Based on the fluxes of our candidates, we would expect $\sim 50$ sources across the COSMOS Web survey area bright enough to be detected across all four observational filters. 
\end{enumerate}

These sources provide an exciting opportunity for follow-up observations with targeted JWST NIRSpec spectroscopy, considering their positions within the Milky Way. A JWST NIRSpec + MIRI spectrum at 1 - 21$\mu$m of a T$_{\mathrm{eff}} = 467$K Y-dwarf was recently presented in \citet{beiler2023}. This source presented a challenge to theoretical models, demonstrating the importance of increasing the number of these sources explored spectroscopically. Similarly, NIRSpec observations for three brown dwarfs found in the UNCOVER data \citep{langeroodi2023, burgasser2023} were used to understand the atmospheres of these distant sources. Both sets of authors found evidence which suggests these sources are in the thick disk or halo. The existence of sources with these colors should be considered when selecting samples of high-redshift galaxies with signatures of black hole accretion. 

This study demonstrates the importance of the NIRCam wavelength range of 2 - 5$\mu$m in selecting very faint, low-mass brown dwarfs. While extragalactic deep fields are designed to point outside the plane of our Galaxy to minimize stellar contamination, the usage of JWST deep observations to collect these sources has implications for the population of objects outside of the thick disk, which may be among the most metal-poor and oldest at these temperatures in the Galaxy. We expect that the large bounty of brown dwarf candidates found over the next several years will shed light on the history of star and planet formation in our Galaxy. 

\begin{acknowledgments}
The authors want to thank the anonymous referee for their comments which greatly improved this manuscript. Additionally, the authors wish to thank Mark Marley for helpful discussions of brown dwarf atmospheric modeling. This work is based on observations made with the NASA/ESA/CSA James Webb Space Telescope, as part of JADES \citep{JADES} and CEERS \citep{CEERS}. The data were obtained from the Mikulski Archive for Space Telescopes at the Space Telescope Science Institute, which is operated by the Association of Universities for Research in Astronomy, Inc., under NASA contract NAS5-03127 for JWST. These observations are associated with PID 1063, 1345, 1180, 1181, 1210, 1286, 1963, 1837, 1895, and 2738. Additionally, this work made use of the lux supercomputer at UC Santa Cruz which is funded by NSF MRI grant AST1828315, as well as the High Performance Computing (HPC) resources at the University of Arizona which is funded by the Office of Research Discovery and Innovation (ORDI), Chief Information Officer (CIO), and University Information Technology Services (UITS). We acknowledge support from the NIRCam Science Team contract to the University of Arizona, NAS5-02015. Funding for this research was provided by the Johns Hopkins University, Institute for Data Intensive Engineering and Science (IDIES). W.B. acknowledges support by the Science and Technology Facilities Council (STFC), ERC Advanced Grant 695671 "QUENCH". REH acknowledges acknowledges support from the National Science Foundation Graduate Research Fellowship Program under Grant No. DGE-1746060. The research of CCW is supported by NOIRLab, which is managed by the Association of Universities for Research in Astronomy (AURA) under a cooperative agreement with the National Science Foundation. The authors are extremely grateful to the CEERS team lead by PI Steve Finkelstein and the FRESCO team led by PI Pascal Oesch for their work developing their observing programs and the zero-exclusive-access period on both. The authors also acknowledge Anton Koekemoer for the usage of CEERS HST data. 
\end{acknowledgments}

\vspace{5mm}
\facilities{JWST(NIRCam, NIRSpec), HST(ACS, WFC3), Spitzer(IRAC)}
\software{{\tt astropy} \citep{astropy2013,astropy2018,astropy2022}, {\tt astroquery} \citep{gisburg2019}, {\tt matplotlib} \citep{matplotlib2007}, {\tt numpy} \citep{numpy2020}, {\tt scipy} \citep{scipy2020}, {\tt Photutils}  \citep{photutils2023}, {\tt EAZY} \citep{brammer2008}. 
          }

\bibliography{main}{}
\bibliographystyle{aasjournal}

\end{document}